\documentclass[lettersize,journal]{IEEEtran}
\usepackage{amsmath,amsfonts}
\usepackage{algorithmic}
\usepackage{algorithm}
\usepackage{array}
\usepackage[caption=false,font=normalsize,labelfont=sf,textfont=sf]{subfig}
\usepackage{textcomp}
\usepackage{stfloats}
\usepackage{url}
\usepackage{verbatim}
\usepackage{graphicx}
\usepackage{cite}
\usepackage{svg}
\begin{document}

\title{PointEMRay: A Novel Efficient SBR Framework on Point Based Geometry}

\author{Kaiqiao Yang, Che Liu ~\IEEEmembership{Member,~IEEE}, Wenming Yu, and Tie Jun Cui ~\IEEEmembership{Fellow,~IEEE}
\thanks{This work was supported by National Natural Science Foundation of China (No.62301146), Natural Science Foundation of Jiangsu Province of China (No.BK20230816) and China Postdoctoral Science Foundation (No.2023M730554 and No.BX20220065). Kaiqiao Yang and Che Liu contributed equally to this work. Kaiqiao Yang and Che Liu designed the  relevant algorithms and prepared the manuscript. Kaiqiao Yang wrote the main codes. Tie Jun Cui and Wenming Yu initiated and supervised the research. All authors contributed to the data analysis and writing of the manuscript, which was reviewed by all authors. The corresponding author: Che Liu.}
\thanks{Kaiqiao Yang, Che Liu, Wenming Yu and Tie Jun Cui are with the Institute of Electromagnetic Space, and the State Key Laboratory of Millimeter Wave, Southeast University, Nanjing 211189, China, and Pazhou Lab. (Huangpu), Guangzhou, China. (e-mail: cheliu@seu.edu.cn).}}

\markboth{Journal of \LaTeX\ Class Files,~Vol.~14, No.~8, August~2021}%
{Shell \MakeLowercase{\textit{et al.}}: A Sample Article Using IEEEtran.cls for IEEE Journals}


\maketitle

\begin{abstract}
The rapid computation of electromagnetic (EM) fields across various scenarios has long been a challenge, primarily due to the need for precise geometric models. The emergence of point cloud data offers a potential solution to this issue. However, the lack of electromagnetic simulation algorithms optimized for point-based models remains a significant limitation. In this study, we propose PointEMRay, an innovative shooting and bouncing ray (SBR) framework designed explicitly for point-based geometries. To enable SBR on point clouds, we address two critical challenges: point-ray intersection (PRI) and multiple bounce computation (MBC). For PRI, we propose a screen-based method leveraging deep learning. Initially, we obtain coarse depth maps through ray tube tracing, which are then transformed by a neural network into dense depth maps, normal maps, and intersection masks, collectively referred to as geometric frame buffers (GFBs). For MBC, inspired by simultaneous localization and mapping (SLAM) techniques, we introduce a GFB-assisted approach. This involves aggregating GFBs from various observation angles and integrating them to recover the complete geometry. Subsequently, a ray tracing algorithm is applied to these GFBs to compute the scattering electromagnetic field. Numerical experiments demonstrate the superior performance of PointEMRay in terms of both accuracy and efficiency, including support for real-time simulation. To the best of our knowledge, this study represents the first attempt to develop an SBR framework specifically tailored for point-based models.
\end{abstract}


\section{Introduction}
\IEEEPARstart{T}{he} shooting and bouncing ray (SBR) technique is a well-established high-frequency asymptotic method (HFAM) in computational electromagnetics (CEM), specifically designed for large-scale electromagnetic problems. Renowned for its remarkable efficiency and accuracy, SBR has been extensively employed in the analysis of scattering characteristics \cite{ref1, ref2, ref3}, channel modeling and measurements \cite{ref4, ref5, ref6}, and electromagnetic imaging \cite{ref7, ref8, ref9}. However, most contemporary SBR simulators rely on mesh-based geometry representations, which are manageable for simple scenes or pre-existing meshes but pose substantial efficiency challenges in complex environments or regions where generating high-quality meshes is quite difficult. Therefore, integrating more versatile and accessible geometric representations into the SBR computational framework is imperative to enhance its applicability in diverse scenarios.

Point cloud is one of the most popular 3D modeling techniques, which samples massive points directly on the surface of an object and provides a simple geometry representation. Today, many advanced devices have been designed to powerfully accelerate the collection of point clouds, such as LiDAR \cite{ref13} and RGB-D cameras \cite{ref14}. Consequently, these strengths make point clouds much more suitable than meshes for fast modeling and simulation. Up till now, some studies have already been conducted on the possible applications of point clouds in CEM \cite{ref10, ref11, ref12, ref35, ref36, ref38, ref39, ref40}.

This paper primarily focuses on developing a general point cloud-based SBR framework that can achieve accuracy and efficiency comparable to mesh-based methods. Since both rays and points are "singular" primitives \cite{ref20}, most solutions involve either tracing ray "tubes" (e.g., cylinders, cones, and even ellipsoids) \cite{ref12, ref19, ref35} or assigning finite surface areas to points \cite{ref15, ref20, ref21}. We categorize these solutions into two main types: screen-based point-ray intersection methods (SPRIMs) \cite{ref12, ref19, ref35} and model-based point-ray intersection methods (MPRIMs) \cite{ref15, ref16, ref17, ref18, ref20, ref21, ref22}. MPRIMs can be further divided into mesh-based (MB-MPRIMs) \cite{ref15, ref16, ref17, ref18} and mesh-free (MF-MPRIMs) \cite{ref20, ref21, ref22} approaches, based on whether a triangular mesh is reconstructed. Additionally, with advancements in artificial intelligence, both SPRIMs and MPRIMs begin leveraging intelligent technologies to enhance computational performance \cite{ref23, ref24, ref25, ref26, ref27, ref30}, leading to some remarkable recent developments.

SPRIM starts by assigning a finite volume to rays. Because rays are typically generated from the "screen" of wavefront, these methods are referred to as screen-based. In SPRIMs, a ray tube is launched from the screen to the point cloud, fetching a set of points, blended to extract desired geometrical features, such as depth and normal vectors \cite{ref19}. Järveläinen et al. used ellipsoid models to simulate indoor wireless channels based on LiDAR point clouds \cite{ref12, ref35}. Recently in computer graphics (CG), Chang et al. introduced neural networks to improve blending performance, generating high-quality images \cite{ref30}. Although these studies \cite{ref12, ref35, ref30} successfully addressed multi-path effect simulation, they all rely on regularly distributed points. Consequently, careful data preprocessing is necessary to avoid holes and artifacts. To overcome this issue, it may be a good choice to introduce wavefront "filters" to repair defects \cite{ref28, ref29}, as some recent inverse EM scattering researches. While the filter-assisted SPRIM may be highly efficient, it will fail to support multi-bounce ray tracing due to the lack of complete 3D information. 

MB-MPRIMs offer a direct solution of reconstructing triangle meshes by many pratical algorithms, including Delaunay triangulation \cite{ref16}, alpha shape reconstruction \cite{ref17}, and Poisson surface reconstruction \cite{ref18}. In the recent years, MB-MPRIMs are also widely used in the area of CEM. Pang et al. adopted region growing algorithm to recover walls from given point cloud \cite{ref15}. In \cite{ref31}, Poisson reconstruction and ray tracing were used to investigate wireless channel in railway tunnels. Zhang et al. employed plane fitting and Poisson reconstruction to recover 3D buildings from drone aerial photos \cite{ref32}. And In \cite{ref36}, reconstruction method was used in industrial scenario simulations. In CG, neural networks are recently used for direct reconstructions \cite{ref23, ref24}. However, these methods often struggle with handling holes, sharp corners, and point noise, and usually have high time complexity. Moreover, the selection of hyperparameters \cite{ref17, ref32} can significantly impact mesh reconstruction quality. While finely estimated point normal vectors have been shown to enhance the quality of mesh reconstruction \cite{ref25, ref33, ref34}, however, these methods still remain time-consuming.

Instead of restoring strict connections, MF-MPRIMs treat point clouds more flexibly. These methods typically assign basis primitives to each point in 3D space, such as splats \cite{ref21, ref37}, voxels \cite{ref22}, etc. In the early stage of CG, some studies achieved high-quality global illumination, including multi-bounce reflection and refraction, using these methods \cite{ref21, ref37}. These have inspired many CEM studies. For example, in \cite{ref15} small patches are used for corner recovery, and in \cite{ref39, ref40} splats are adopted in the traditional ray tracing pipeline. However, due to the non-compact tiling of these primitives, most MF-MPRIMs require a massive number of point samples (usually millions \cite{ref37}) and careful hyperparameter selection to avoid holes and artifacts \cite{ref15, ref21, ref37, ref39, ref40}. In \cite{ref38}, Saito et al. utilized sparse points denote scatterers, however, this is not suitable for precise simulations. Recently, the emergence of NeRF \cite{ref26} and Gaussian splatting \cite{ref27} have offered new ways to generate dense geometry representations without extensive hyperparameter tuning. But they require per-scene training and usually have difficulty supporting global illumination. Moreover, the large storage requirements can be a significant challenge.

In this paper, we propose PointEMRay, a general and efficient SBR framework for analyzing electromagnetic (EM) scattering characteristics directly on point cloud represented perfect electric conductor (PEC) targets. For simplicity, we consider only plane wave excitation and neglect diffraction effects. To perform effective SBR on point clouds, we address two primary problems, namely point-ray intersection (PRI) and multiple bounce computation (MBC). For PRI, we adopt the ideas from SPRIMs \cite{ref28, ref29, ref30}, combining ray tube tracing with a convolutional neural network (CNN) to determine the correct intersection position. Compared with MPRIMs \cite{ref18, ref21, ref33}, our screen-based method is accurate and time-efficient, enabling fast single bounce SBR, also known as the physical optics (PO) method. To solve the MBC problem, we need sufficient 3D information. Inspired by simultaneous localization and mapping (SLAM) techniques \cite{ref42, ref43, ref44}, we store dense geometry in several geometric frame buffers (GFBs) generated by the CNN, including depth and normal maps and then trace rays on these GFBs with multiple bounces to calculate the scattering field. Experiments demonstrate that PointEMRay achieves excellent accuracy and efficiency compared to several traditional methods. Additionally, we implemented parallel acceleration for PointEMRay using GPUs, enabling real-time simulation. To the best of our knowledge, PointEMRay is the first point geometry-based SBR framework designed for EM scattering characteristics analysis.

This paper is structured as follows: Section II introduces the mathematical framework of the SBR method. Section III provides a detailed description of PointEMRay, including the screen-based PRI technique and GFB assisted MBC. In Section IV, we present numerical results that demonstrate the accuracy and efficiency of PointEMRay. Finally, Section V summarizes the conclusions drawn from our findings and offers perspectives for future research.

\section{Fundamentals of the SBR Method}

\begin{figure}[!t]
\centering
\includegraphics[width=3.5in]{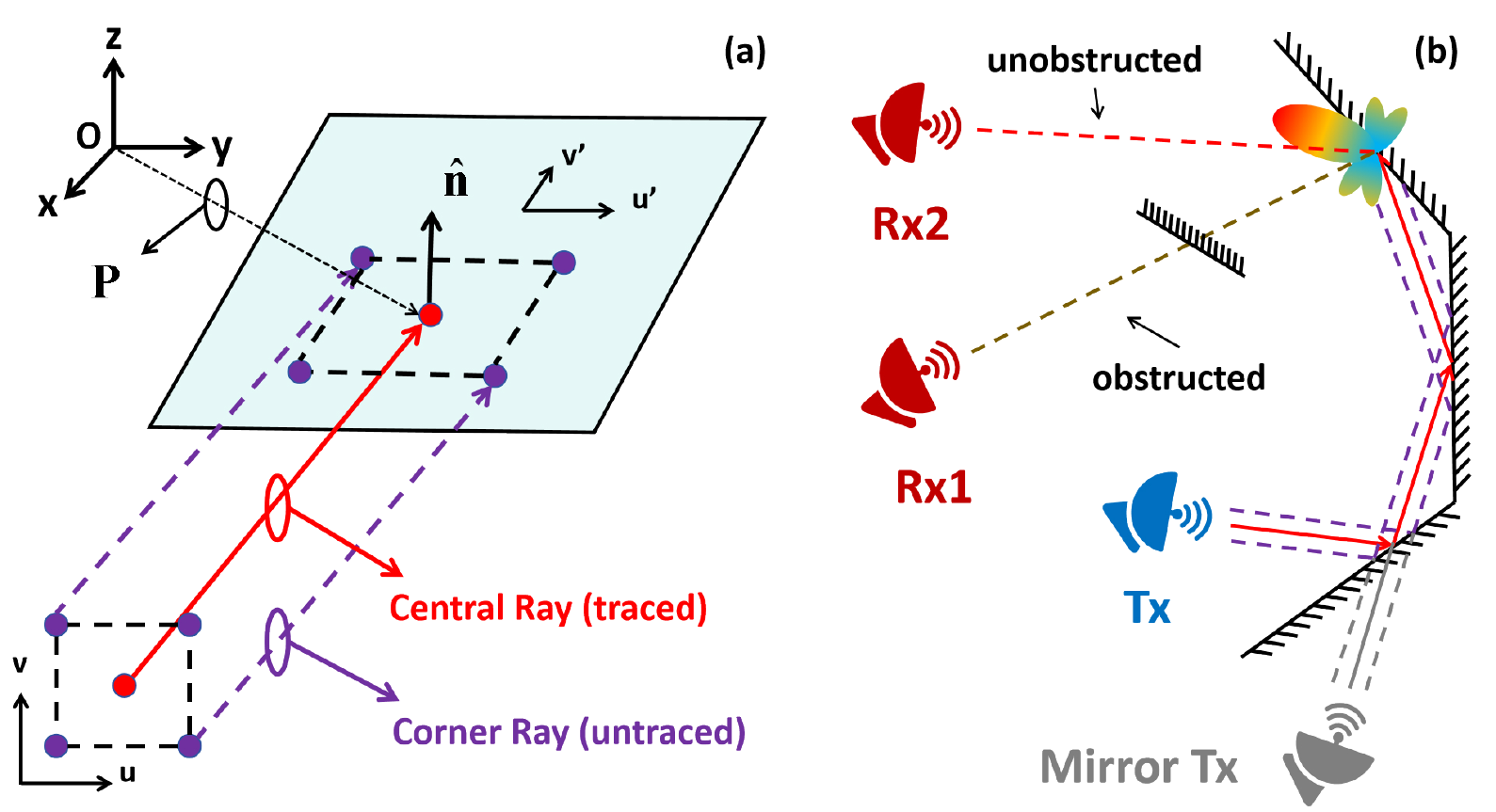}
\caption{Illustration of PO and SBR methods. (a) The PO method where only central rays are traced. (b) The SBR method.}
\label{fig_1}
\end{figure}

\subsection{GO Method}
As the frequency increases, EM waves in free space exhibit characteristics akin to optical rays. Consequently, the same approach can be employed to study the propagation of high-frequency EM waves. Following the methodology outlined in \cite{ref45}, the electric field $\mathbf{E}$ along the ray path can be computed as shown in Equation \eqref{deqn_ex1}:
\begin{equation}
    \label{deqn_ex1}
    \mathbf{E}(s) = \mathbf{E}_{0} \cdot \left(\prod^N_{i=1}\overline{\overline{\mathbf{R}}}_i\right) A(s) e^{-jk_0s+\phi_0}
\end{equation}
\noindent In the equation, $\mathbf{E}_{0}$ represents the incident field's amplitude, including its polarization information. $N$ is the total number of bounces, $\overline{\overline{\mathbf{R}}}$ stands for the dyadic reflection factor, $A$ is the spreading factor, $k_0$ is the wave number in vacuum, and $\phi_0$ signifies the initial phase of the incident field. Given our focus on far-field scattering, we always have $A=1$.

As the reflection happens, polarization adjusts to meet the boundary condition. We denote the reflection and incident fields separately as $\mathbf{E}^r$ and $\mathbf{E}^i$. At the intersection position $s_0$, they relate as follows:
\begin{equation}
    \label{deqn_ex2}
    \mathbf{E}^r(s_0) = \mathbf{E}^i(s_0) \cdot \overline{\overline{\mathbf{R}}}(s_0)
\end{equation}
\noindent To elaborate further, we can express $\mathbf{E}^r$ and $\mathbf{E}^i$ as follows: $\mathbf{E}^r = E^r_p\hat{\mathbf{e}}^r_p + E^r_v\hat{\mathbf{e}}_v$ and $\mathbf{E}^i = E^i_p\hat{\mathbf{e}}^i_p + E^i_v\hat{\mathbf{e}}_v$. Here, $(\cdot)_p$ and $(\cdot)_v$ represent the parallel and vertical polarization components, respectively. We determine $\hat{\mathbf{e}}_v$ as $(\hat{\mathbf{k}}^i\times\hat{\mathbf{n}}) / |\hat{\mathbf{k}}^i\times\hat{\mathbf{n}}|$, $\hat{\mathbf{e}}^r_p$ as $(\hat{\mathbf{e}}_v\times\hat{\mathbf{k}}^r) / |\hat{\mathbf{e}}_v\times\hat{\mathbf{k}}^r|$, and $\hat{\mathbf{e}}^i_p$ as $(\hat{\mathbf{e}}_v\times\hat{\mathbf{k}}^i) / |\hat{\mathbf{e}}_v\times\hat{\mathbf{k}}^i|$, where $\hat{(\cdot)}$ denotes unit vectors, $\mathbf{k}^r$ and $\mathbf{k}^i$ represent the reflection and incident wave vectors, respectively, and $\hat{\mathbf{n}}$ denotes the surface normal vector. Ultimately, we arrive at Equation \eqref{deqn_ex3}:
\begin{equation}
    \label{deqn_ex3}
    \begin{bmatrix} 
        E^r_p(s_0) \\ 
        E^r_v(s_0)  
    \end{bmatrix} = 
    \begin{bmatrix} 
        \Gamma_p(s_0) & 0 \\ 
        0 & \Gamma_v(s_0)  
    \end{bmatrix}
    \begin{bmatrix} 
        E^i_p(s_0) \\ 
        E^i_v(s_0)  
    \end{bmatrix}
\end{equation}
\noindent Here, $\Gamma$ denotes the scalar reflection factor. Applying the PEC constraint, it is straightforward to ascertain that $\Gamma_p(s_0) = 1$ and $\Gamma_v(s_0) = -1$.

When a ray strikes the surface of the target, its propagation direction alters. According to Snell's Law on the PEC surface, the reflection wave vector $\hat{\mathbf{k}}^r$ can be computed as follows:
\begin{equation}
    \label{deqn_ex4}
    \hat{\mathbf{k}}^r = \hat{\mathbf{k}}^i \cdot \left(\overline{\overline{\mathbf{I}}} - 2\hat{\mathbf{n}}\hat{\mathbf{n}}\right)
\end{equation}
\noindent where $\overline{\overline{\mathbf{I}}}$ is the unit dyadic. 

This shooting and bouncing process continues iteratively until the termination conditions are satisfied.

\subsection{PO Method}
The PO method, stemming from wave physics, efficiently computes the continuous scattering field within the observation interval. Utilizing the Kirchhoff diffraction integral under far-field and PEC assumptions \cite{ref45}, we obtain the scattering field $\mathbf{E}^{s}$:
\begin{equation}
    \label{deqn_ex5}
    \begin{split}
        \mathbf{E}^{s}(\mathbf{r}) &= -\frac{j\eta_0k_0}{2 \pi} \cdot \\
        &\iint_{S'} \left[ \left( \hat{\mathbf{n}} \times \mathbf{H}^{i} \right) \times \hat{\mathbf{k}}^{s} \right] \times \hat{\mathbf{k}}^{s} e^{j\mathbf{k}^{s} \cdot \mathbf{r}'} dS'
    \end{split}
\end{equation}
\noindent Here, $\eta_0$ represents vacuum wave impedance, $\mathbf{H}^{i}$ denotes incident magnetic field, and $\mathbf{k}^{s}$ stands for the scattering wave vector. Substituting $\mathbf{H}^{i} = \hat{\mathbf{h}} e^{-j\mathbf{k}^i \cdot \mathbf{r}'}$ yields Equation \eqref{deqn_ex6}:
\begin{equation}
    \label{deqn_ex6}
    \begin{split}
        \mathbf{E}^{s}(\mathbf{r}) &= -\frac{j\eta_0k_0}{2 \pi} \cdot \\
        &\iint_{S'} \left[ \left( \hat{\mathbf{n}} \times \hat{\mathbf{h}} \right) \times \hat{\mathbf{k}}^{s} \right] \times \hat{\mathbf{k}}^{s} e^{-j\left(\mathbf{k}^{i} - \mathbf{k}^{s} \right) \cdot \mathbf{r}'} dS'
    \end{split}
\end{equation}
Here, $\mathbf{h}$ represents the polarization vector of the incident magnetic field, and $\mathbf{k}^{i}$ denotes the incident wave vector. According to \cite{ref46}, Equation \eqref{deqn_ex6} has an analytical expression if the illuminated area $S'$ consists of polygons.

In traditional PO frameworks, the tracing procedure is typically divided into corner ray tracing and central ray tracing stages, as shown in \cite{ref47}, to form the ray tube structure. However, this approach is redundant as central rays already provide adequate information when evenly distributed.

Equation \eqref{deqn_ex6} can be alternatively expressed as \eqref{deqn_ex7}:
\begin{equation}
    \label{deqn_ex7}
    \begin{split}
        \mathbf{E}^{s}(\mathbf{r}) &= -\frac{j\eta_0k_0}{2 \pi} \cdot \\
        &\left[ \left( \hat{\mathbf{n}} \times \hat{\mathbf{h}} \right) \times \hat{\mathbf{k}}^{s} \right] \times \hat{\mathbf{k}}^{s} \iint_{S'} e^{-j\left(\mathbf{k}^{i} - \mathbf{k}^{s} \right) \cdot \mathbf{r}'} dS'
    \end{split}
\end{equation}
\noindent We denote the integral term in Equation \eqref{deqn_ex7} as $I$ and decompose the local position $\mathbf{r}'$ on the illuminated surface as $\mathbf{r}'=\mathbf{p}+\mathbf{x}'$, where $\mathbf{p}$ is the intersection position of a central ray in Fig. 1. Thus, we obtain Equation \eqref{deqn_ex8}: 
\begin{equation}
    \label{deqn_ex8}
    I = e^{-j\left( \mathbf{k}^i - \mathbf{k}^s \right) \cdot \mathbf{p}} \iint_{S'} e^{-j\left( \mathbf{k}^i - \mathbf{k}^s \right) \cdot \mathbf{x}'} dS'
\end{equation}

Since the incident EM rays are parallel, the corner rays can collectively define a tangent plane of the parallelogram. As depicted in Fig. 1(a), vector $\mathbf{x}'$ can be approximated as $\mathbf{x}'\approx u'\hat{\mathbf{u}}'+v'\hat{\mathbf{v}}'$. Consequently, we derive Equation \eqref{deqn_ex9}:
\begin{equation}
    \label{deqn_ex9}
    \begin{split}
        I &\approx \\&e^{-j\left( \mathbf{k}^i - \mathbf{k}^s \right) \cdot \mathbf{p}} \int_{-\frac{U'}{2}}^{\frac{U'}{2}} \int_{-\frac{V'}{2}}^{\frac{V'}{2}} e^{-j\left( \mathbf{k}^i - \mathbf{k}^s \right) \cdot \left( u'\hat{\mathbf{u}}'+v'\hat{\mathbf{v}} \right)} dv'du'  
    \end{split}
\end{equation}
\noindent where $U'$ and $V'$ are the lengths of vector $\mathbf{u}'$ and $\mathbf{v}'$. The double integral in \eqref{deqn_ex9} can be separated, yielding \eqref{deqn_ex10}
\begin{equation}
    \label{deqn_ex10}
    \begin{split}
        I &\approx \\
        &\int_{-\frac{U'}{2}}^{\frac{U'}{2}} e^{-j\left( \mathbf{k}^i - \mathbf{k}^s \right) \cdot u'\hat{\mathbf{u}}'} du' \cdot \int_{-\frac{V'}{2}}^{\frac{V'}{2}} e^{-j\left( \mathbf{k}^i - \mathbf{k}^s \right) \cdot v'\hat{\mathbf{v}}} dv' \cdot \\
        &e^{-j\left( \mathbf{k}^i - \mathbf{k}^s \right) \cdot \mathbf{p}}
    \end{split}
\end{equation}
\noindent and then \eqref{deqn_ex11}
\begin{equation}
    \label{deqn_ex11}
    \begin{split}
        I &\approx \\
        &U'V' \mathrm{sinc}\left[ \frac{\left( \mathbf{k}^i - \mathbf{k}^s \right) \cdot \mathbf{u}'}{2} \right] \mathrm{sinc}\left[ \frac{\left( \mathbf{k}^i - \mathbf{k}^s \right) \cdot \mathbf{v}'}{2} \right] \cdot \\
        &e^{-j\left( \mathbf{k}^i - \mathbf{k}^s \right) \cdot \mathbf{p}} 
    \end{split}
\end{equation}

Previously, we derived these expressions using local vectors $\mathbf{u}'$ and $\mathbf{v}'$. Now, we can readily transform Equation \eqref{deqn_ex11} into the screen or antenna coordinate system using Equations \eqref{deqn_ex12} and \eqref{deqn_ex13}:
\begin{equation}
    \label{deqn_ex12}
    \mathbf{u}'=\mathbf{u}-\hat{\mathbf{k}}^i\frac{\hat{\mathbf{n}} \cdot \mathbf{u}}{\hat{\mathbf{n}} \cdot \hat{\mathbf{k}}^i}
\end{equation}
\begin{equation}
    \label{deqn_ex13}
    \mathbf{v}'=\mathbf{v}-\hat{\mathbf{k}}^i\frac{\hat{\mathbf{n}} \cdot \mathbf{v}}{\hat{\mathbf{n}} \cdot \hat{\mathbf{k}}^i}
\end{equation}

To compute the integral $I$, the remaining unknown parameters are solely the intersection position $\mathbf{p}$ and the surface normal $\hat{\mathbf{n}}$, both of which are exclusively associated with the central ray. This streamlined approach reduces computational overhead and facilitates alignment of the GFBs.

\subsection{Field Calculation}
To calculate the total field, we aggregate contributions from all EM rays. Each EM ray undergoes initial processing using the GO method to determine its final intersection position and corresponding normal vector. Subsequently, we assess visibility to determine if any obstructions lie along the line-of-sight between the receiver and the final intersection position. Finally, we employ the PO method to calculate the field contribution of each specific ray.

Fig. 1(b) depicts the process of a complete SBR procedure. Here, we utilize the image method, as demonstrated in \cite{ref45}, to determine the distance $d$ traveled by an EM ray. Finally, we establish the comprehensive scattering field calculation formula for the SBR method, represented by Equation \eqref{deqn_ex14}. In this equation, $l$ denotes the $l$-th EM ray, $N$ represents the total number of EM rays, $M_l$ signifies the maximum bounce of the $l$-th EM ray, $\Phi$ indicates the validity of an EM ray ($\Phi=1$ if the ray hits a target, otherwise $\Phi=0$), $\Psi$ reflects the visibility of the receiver for the last intersection position ($\Psi=1$ if the receiver is unobstructed, otherwise $\Psi=0$), and $\mathbf{\Theta}^{(0)}$ denotes transmitter parameters. 

Equation \eqref{deqn_ex14} delves into a pivotal consideration: the computational accuracy of HFAM is markedly impacted by geometric intricacies, including the orientation of normal vectors and intersection positions. This realization will serve as a cornerstone for the advancement of the PointEMRay framework.
\begin{figure*}
    \begin{equation}
        \label{deqn_ex14}
        \begin{split}
            \mathbf{E}^s&=-\frac{j\eta_0k_0}{2 \pi}\sum_{l=1}^N \Phi_l \Psi_l^{(M_l)} U'^{(M_l)}_lV'^{(M_l)}_l \mathrm{sinc}\left[ \frac{\left( \mathbf{k}^{i, (M_l - 1)}_l - \mathbf{k}^s \right) \cdot \mathbf{u}'^{(M_l)}_l}{2} \right] \mathrm{sinc}\left[ \frac{\left( \mathbf{k}^{i, (M_l - 1)}_l - \mathbf{k}^s \right) \cdot \mathbf{v}'^{(M_l)}_l}{2} \right] \cdot \\
            &\left[ \left( \hat{\mathbf{n}}_l^{(M_l)} \times \hat{\mathbf{h}}^{(M_l - 1)}_l \right) \times \hat{\mathbf{k}}^{s} \right] \times \hat{\mathbf{k}}^{s} e^{-j\left( k_0d^{(M_l)}_l - \mathbf{k}^s \cdot \mathbf{p}^{(M_l)}_l \right)} \\
            &=\sum_{l=1}^N\Phi_l\mathbf{A}\left( \hat{\mathbf{n}}^{(1)}_l, \hat{\mathbf{n}}^{(2)}_l, ..., \hat{\mathbf{n}}^{(M_l)}_l; \mathbf{\Theta^{(0)}} \right) e^{j\varphi \left( \hat{\mathbf{n}}^{(1)}_l, \hat{\mathbf{n}}^{(2)}_l, ..., \hat{\mathbf{n}}^{(M_l)}_l, d_l^{(M_l)}; \mathbf{\Theta}^{(0)} \right)}
        \end{split}
    \end{equation}
\end{figure*}

\section{PointEMRay Architecture}
This section offers an insight into the architecture of the PointEMRay framework, illustrated in Fig. 2. We commence with an introduction to our screen-based PRI process, responsible for accurate intersection prediction and GFB generation, which is essential for following computations. Subsequently, we delve into the methodology for executing MBC utilizing the pre-established GFBs. Finally, we explore the time complexity inherent in the PointEMRay framework.

\begin{figure*}[!t]
\centering
\includegraphics[width=7.0in]{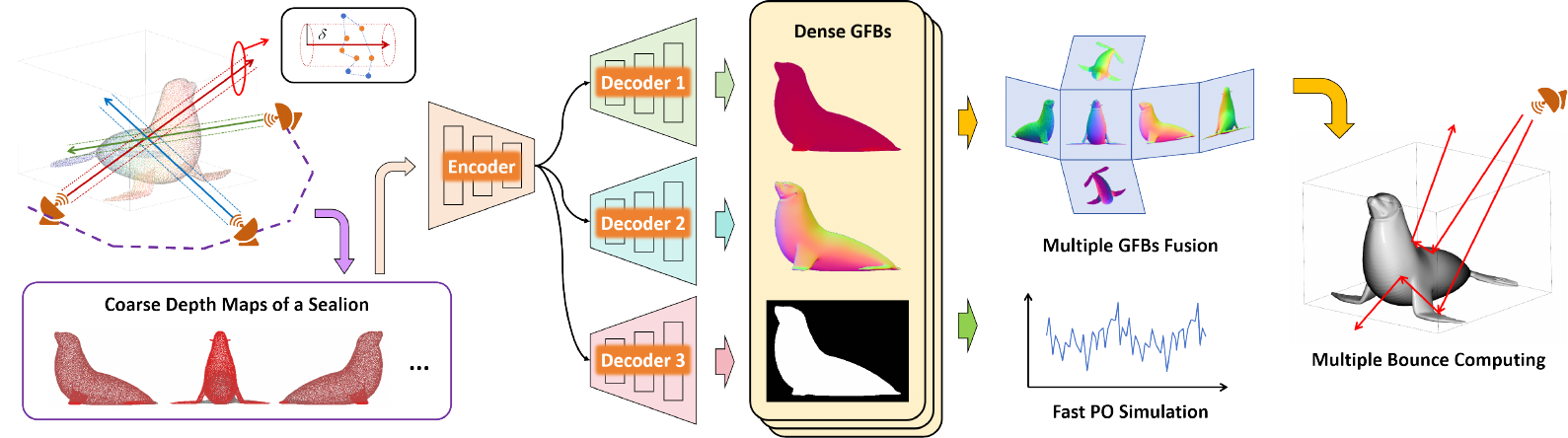}
\caption{The PointEMRay framework initiates with a raw point cloud input and proceeds to simulate the corresponding EM scattering field. To pinpoint intersections on the point cloud, PointEMRay adopts a screen-based approach. It begins by projecting a multitude of cylindrical rays to generate coarse depth maps. Subsequently, a neural network refines these maps, producing corresponding oriented normal maps and intersection masks. These datasets collectively constitute sets of GFBs. Depending on the application, users can choose between a swift PO simulation or execute multiple GFB fusions to compute the multi-path effects.}
\label{fig_2}
\end{figure*}

\subsection{Screen-Based PRI}
In point-based ray methods, PRI stands out as a crucial challenge. Given that both points and rays are singular primitives, densifying the 3D representation and pinpointing potential intersection positions necessitates assigning finite surface areas to either. While MPRIMs offer robust tools for finely recovering target surfaces, they grapple with issues such as high time complexity, sensitivity to hyperparameters, and difficulty handling sharp structures in practical scenarios. Here, we turn to SPRIMs for assistance. To estimate intersection points on the target surface, we amalgamate a ray tube tracing method with a post-processing neural network, as illustrated in Fig. 2. This approach enables us to initially obtain a series of coarse estimations, subsequently refined through neural network corrections. The complete algorithm for our screen-based PRI is outlined in Algorithm \ref{alg:alg1}.

\begin{algorithm}[H]
\caption{Screen Based PRI}\label{alg:alg1}
    \renewcommand{\algorithmicrequire}{\textbf{Input:}}
    \renewcommand{\algorithmicensure}{\textbf{Output:}}
    \begin{algorithmic}[1]
        \REQUIRE $\mathbf{P}$, $sen$, $K$  
        \ENSURE $gfb$   
        
        \STATE $AS\gets$ ASConstruction($\mathbf{P}$)
        \STATE $tubes\gets$ RayTubeInitialization($sen$)

        \FOR{each $tube$ in $tubes$}
            \STATE Allocate $K$ points space to $heap$
            \STATE TraverseASRayTube($tube$, $AS.root$, $pointList$)
            \WHILE{$curPoint\gets pointList$}
                \STATE Heapify($heap$, $curPoint$, $K$)
            \ENDWHILE
            \STATE Min-Heapify($heap$, $K$)
            \STATE $\mathbf{p}\gets heap[0]$
            \STATE $\mathbf{o}\gets tube.org$
            \STATE $tube.dis\gets$ Length($\mathbf{p} - \mathbf{o}$)
        \ENDFOR

        \STATE $\bar{\mathbf{D}} \gets$ GetDepthMap($tubes$)
        \STATE $\tilde{\mathbf{D}}, \mathbf{N}, \mathbf{\Phi}\gets U(\bar{\mathbf{D}}; \theta)$
        \STATE $gfb\gets \{\tilde{\mathbf{D}}, \mathbf{N}, \mathbf{\Phi}\}$
    \end{algorithmic}
\label{alg1}
\end{algorithm}

\begin{algorithm}[H]
\caption{Ray Tube Traversal of AS}\label{alg:alg2}
    \renewcommand{\algorithmicrequire}{\textbf{Input:}}
    \renewcommand{\algorithmicensure}{\textbf{Output:}}
    \begin{algorithmic}[1]
        \REQUIRE $tube$, $node$, $PL$  
        \ENSURE None   
        
        \IF{$tube$ intersects $node$}
            \IF{$node.isLeaf$}
                \STATE $PL$.append(points inside $tube$)
                \STATE return
            \ELSE
                \STATE TraverseASRayTube($tube$, $node.leftChild$, $PL$)
                \STATE TraverseASRayTube($tube$, $node.rightChild$, $PL$)
            \ENDIF
        \ELSE
            \STATE return
        \ENDIF
    \end{algorithmic}
\label{alg2}
\end{algorithm}

We designate the given point set as $\mathbf{P}$. Initially, we emit numerous cylindrical ray tubes from the transmitter (referred to as $sen$ in Algorithm \ref{alg:alg1}, storing both extrinsic and intrinsic parameters) towards $\mathbf{P}$ to identify a subset of $K$ nearest points to the central ray within each ray tube. These points are then processed further to pinpoint the exact intersection position. A hyperparameter $\delta$ defines the radius of our ray tube. Practically, a point becomes a candidate only if its distance to the central ray of the ray tube is less than $\delta$. We adopt the Top-K algorithm's concept of tracing and heapifying. Upon traversing $\mathbf{P}$, we subsequently perform a min-heapify operation based on the distances of $K$ candidate points to the origin of the ray tube. This operation assists in selecting the point closest to the origin of the ray tube, likely the nearest neighbor to the exact intersection position. To expedite the point traversal process, we leverage both acceleration structures (AS) \cite{ref48} and GPU parallel computing techniques to enhance computational efficiency. Subsequently, we compute the distance between the selected point and the origin of the ray tube, forming an initial coarse depth map $\bar{\mathbf{D}}$ of the current viewport. 

Similar to approaches used in depth completion problems \cite{ref29, ref41}, we employ a neural network ($U$ in Algorithm \ref{alg:alg1} with learnable parameters $\theta$) to refine the coarse depth map $\bar{\mathbf{D}}$ into a more detailed one $\tilde{\mathbf{D}}$, while simultaneously predicting the corresponding oriented normal map $\mathbf{N}$ and intersection mask $\mathbf{\Phi}$. Notably, we represent the position of the selected point using depth rather than a 3D $\mathbf{x}\mathbf{y}\mathbf{z}$ vector. This choice is motivated by the fact that global coordinates remain consistent across various observation angles, posing challenges in learning their mapping. We construct our neural network based on the U-Net architecture, as depicted in Fig. 2. Instead of conventional convolution blocks, we employ ConvNeXt blocks \cite{ref49} (illustrated in Fig. 3(a)) as the fundamental elements of the encoder. ConvNeXt has been demonstrated to be lightweight yet powerful, akin to the Transformer architecture \cite{ref50}. For other components of our network, we utilize max-pooling for downsampling and sub-pixel convolution \cite{ref51} for upsampling. Further details regarding the neural network's structure are presented in Fig. 3(b).

\begin{figure}
    \centering
    \includegraphics[width=3.5in]{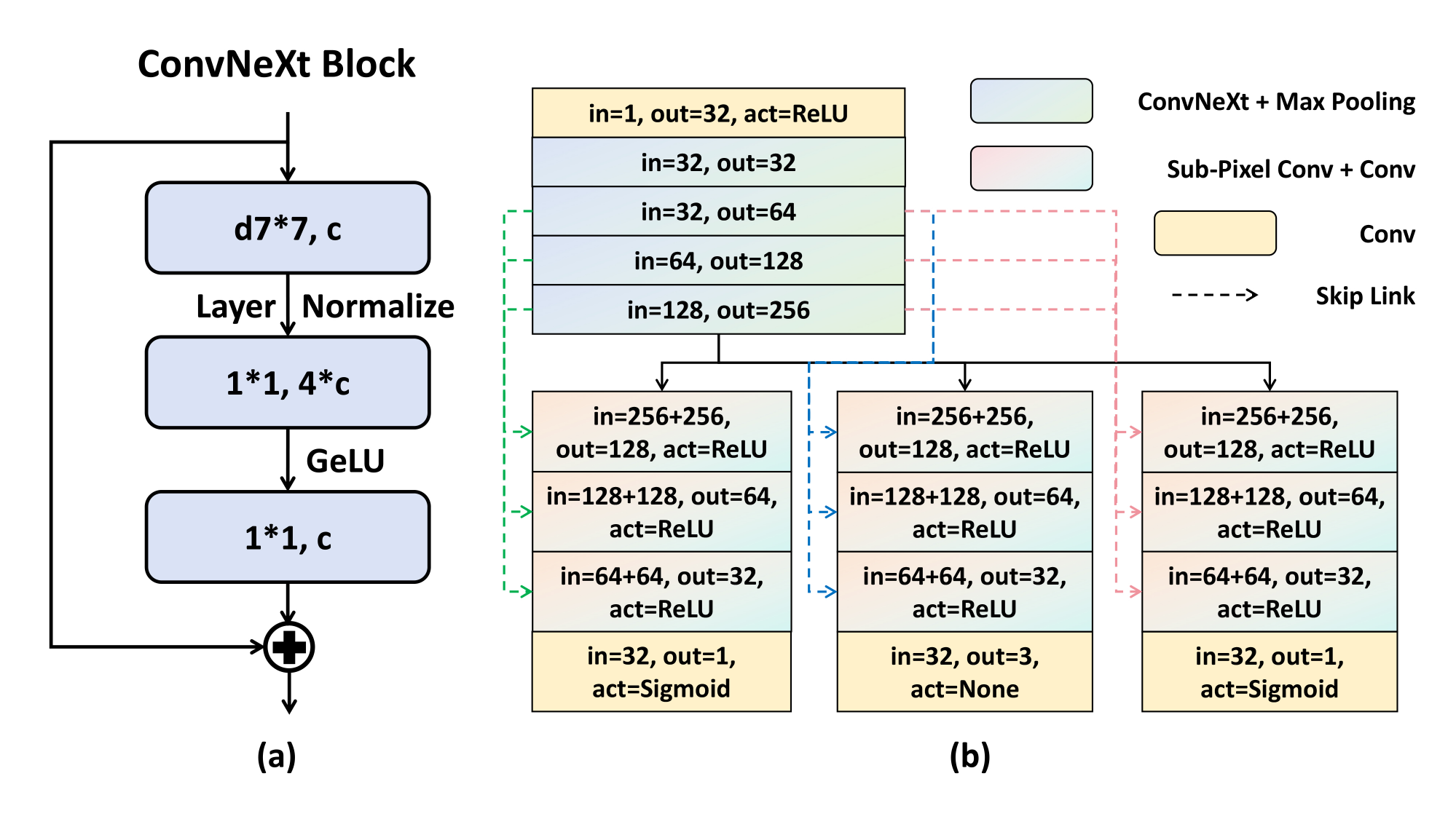}
    \caption{Structures of (a) a ConvNeXt block and (b) our neural network.}
    \label{fig_3}
\end{figure}

We observe that a depth map inherently encapsulates the entire geometric profile of a surface, providing insight into its normal vectors. Specifically, we can establish a local coordinate system $\mathbf{u}\mathbf{v}\mathbf{w}\mathbf{o}$ at the transmitter, where vector $\mathbf{w}$ extends from the target's center to the transmitter. Subsequently, we derive the implicit surface equation as follows:
\begin{equation}
    \label{deqn_ex15}
    F(u, v, w) = w - f(u, v) = 0
\end{equation}
\noindent Assuming we have determined the depth as $d$, we can compute the unit normal vector $\hat{\mathbf{n}}$ at $(u, v, -d)$ as follows:
\begin{equation}
    \label{deqn_ex16}
    \hat{\mathbf{n}} = \frac{\nabla F}{||\nabla F||} = \frac{\mathbf{w} - \dfrac{\partial d}{\partial u} \mathbf{u} - \dfrac{\partial d}{\partial v} \mathbf{v}}{\sqrt{1 + \left(\dfrac{\partial d}{\partial u}\right)^2 + \left(\dfrac{\partial d}{\partial u}\right)^2}}
\end{equation}

\noindent Therefore, our neural network processes the coarse depth map as its sole input, producing a refined depth map, a corresponding normal map, and a hit mask. These outputs are derived from three identical encoders, as illustrated in Fig. 2. 

We trained this neural network using supervised learning. Ground truth depth maps, normal maps, and intersection masks were provided to guide the training process. As previously mentioned, normal maps were transformed into the $\mathbf{u}\mathbf{v}\mathbf{w}\mathbf{o}$ local coordinate system. For the loss function, we first derived equation \eqref{deqn_ex17} to calculate the difference between the predicted depth and the ground truth depth:
\begin{equation}
    \label{deqn_ex17}
    L_1 = ||\tilde{d} - \hat{d}||^2
\end{equation}
\noindent where we denotes the predicted depth map as $\tilde{d}$ and the ground truth map as $\hat{d}$. The depth data is normalized according to the bounding box. Given that orientation is the most critical feature of a normal vector, we use angle differences to describe the deviation of normal vectors, as shown in equation \eqref{deqn_ex18}:
\begin{equation}
    \label{deqn_ex18}
    L_2 = \frac{||\mathbf{n}_p \times \mathbf{n}_g||}{||\mathbf{n}_p||\cdot||\mathbf{n}_g||}
\end{equation}
\noindent where $\mathbf{n}_p$ refers to the predicted normal vector and $\mathbf{n}_g$ refers to the ground truth normal vector. As for the intersection mask, we use the $\alpha$-balanced binary focal loss function \cite{ref52} as 
\begin{equation}
    \label{deqn_ex19}
    L_3 = \begin{cases}
        -\alpha \left( 1-\tilde{h} \right)^\gamma \mathrm{log}\tilde{h}, &{\hat{h}=1} \\
        -(1-\alpha)\tilde{h}^\gamma \mathrm{log\left( 1-\tilde{h} \right)}, &{\hat{h}=0}
    \end{cases}
\end{equation}
\noindent $\tilde{h}$ denotes the predicted intersection probability, while $\hat{h} = {0, 1}$ represents the ground truth intersection probability. We set the hyper-parameters $\alpha$ and $\gamma$ to 0.5 and 2, respectively. Despite these parameters, additional refinement is necessary. Therefore, we incorporate a derivative loss for depth to better capture the detailed geometry. Given that equation \eqref{deqn_ex16} establishes the relationship between a normal vector and the depth derivative, we define the loss function as shown in equation \eqref{deqn_ex20}:
\begin{equation}
    \label{deqn_ex20}
    L_4 = \frac{||\nabla \tilde{d} \times \nabla \hat{d}||}{||\nabla \tilde{d}||\cdot||\nabla \hat{d}||}
\end{equation}
\noindent Finally, we have the total loss function as equation \eqref{deqn_ex21}:
\begin{equation}
    \label{deqn_ex21}
    L = \lambda_1 L_1 + \lambda_2 L_2 + \lambda_3 L_3 + \lambda_4 L_4
\end{equation}
\noindent where $\lambda_1=1$, $\lambda_2=0.5$, $\lambda_3=1$, $\lambda_4=0.5$ are factors.

Currently, screen-based PRI has facilitated the execution of single-bounce SBR or PO algorithms, enabling swift simulations of target objects. Given that many objects exhibit minimal self-reflection, such as vehicles, PO often yields commendable accuracy. To compute the scattering field using the PO method, we simply insert the acquired depth, normal vector, intersection indicator, and other essential variables into equation \eqref{deqn_ex14}. As we solely trace the central rays, the geometry data are inherently aligned, allowing for efficient computation through parallel processing.

\subsection{GFB-Assisted MBC}
In the realm of SPRIMs, addressing MBC has traditionally posed a challenge, particularly when finite volume ray tubes intersect the point cloud in close proximity to the reflection or refraction point. While some studies \cite{ref19, ref35} have assumed a maximum distance between adjacent points, early termination of ray marching remains unavoidable, especially with large reflection angles. One straightforward approach to tackle this issue involves assigning finite surface area to individual points, effectively storing densified 3D information. Splats emerge as an appealing option due to their simplicity and reasonable accuracy. However, determining the optimal radius and normal vector for a splat typically demands densely sampled points and meticulous processing. Inspired by SLAM techniques \cite{ref42, ref43, ref44}, we endeavor to leverage previously generated GFBs to mitigate this challenge. GFBs serve as dense point clouds with evenly distributed points, capturing comprehensive geometric information, including high-quality oriented normal vectors, which were previously computationally intensive to estimate. While each GFB covers only a portion of 3D information from a particular perspective, we preserve a set of them from different perspectives (Fig. 2) to retain the majority of 3D geometric data.

\begin{algorithm}[H]
\caption{GFB Assisted MCB}\label{alg:alg3}
    \renewcommand{\algorithmicrequire}{\textbf{Input:}}
    \renewcommand{\algorithmicensure}{\textbf{Output:}}
    \begin{algorithmic}[1]
        \REQUIRE $gfbs$, $tx$, $rx$, $mb$, $\mathbf{E}^i$  
        \ENSURE $\mathbf{E}_s$   
        
        \STATE $pgfbs\gets$ EdgeFilteringAndFusion($gfbs$)
        \STATE $sgbfs\gets$ ConvertToSplats($pfgs$)
        \STATE $AS\gets$ AsConstruction($sgbfs$)
        \STATE $rays\gets$ RayGenerator($tx$)

        \FOR{$b=1:mb$}
        \FOR{each $ray$ in $rays$}
            \STATE $hitRecord\gets$ TraverseSplatAS($ray$, $AS$)
            \STATE $record[b]$.append($\{hitRecord.dep, hitRecord.pos$,\\                $hitRecord.nor, hitRecord.vis, hitRecord.msk\}$)
        \ENDFOR
        \ENDFOR
        
        \STATE $\mathbf{E}^s\gets$ EMComputing($record, tx, rx, \mathbf{E}^i$)
    \end{algorithmic}
\label{alg3}
\end{algorithm}

\begin{figure}[!t]
\centering
\includegraphics[width=3.5in]{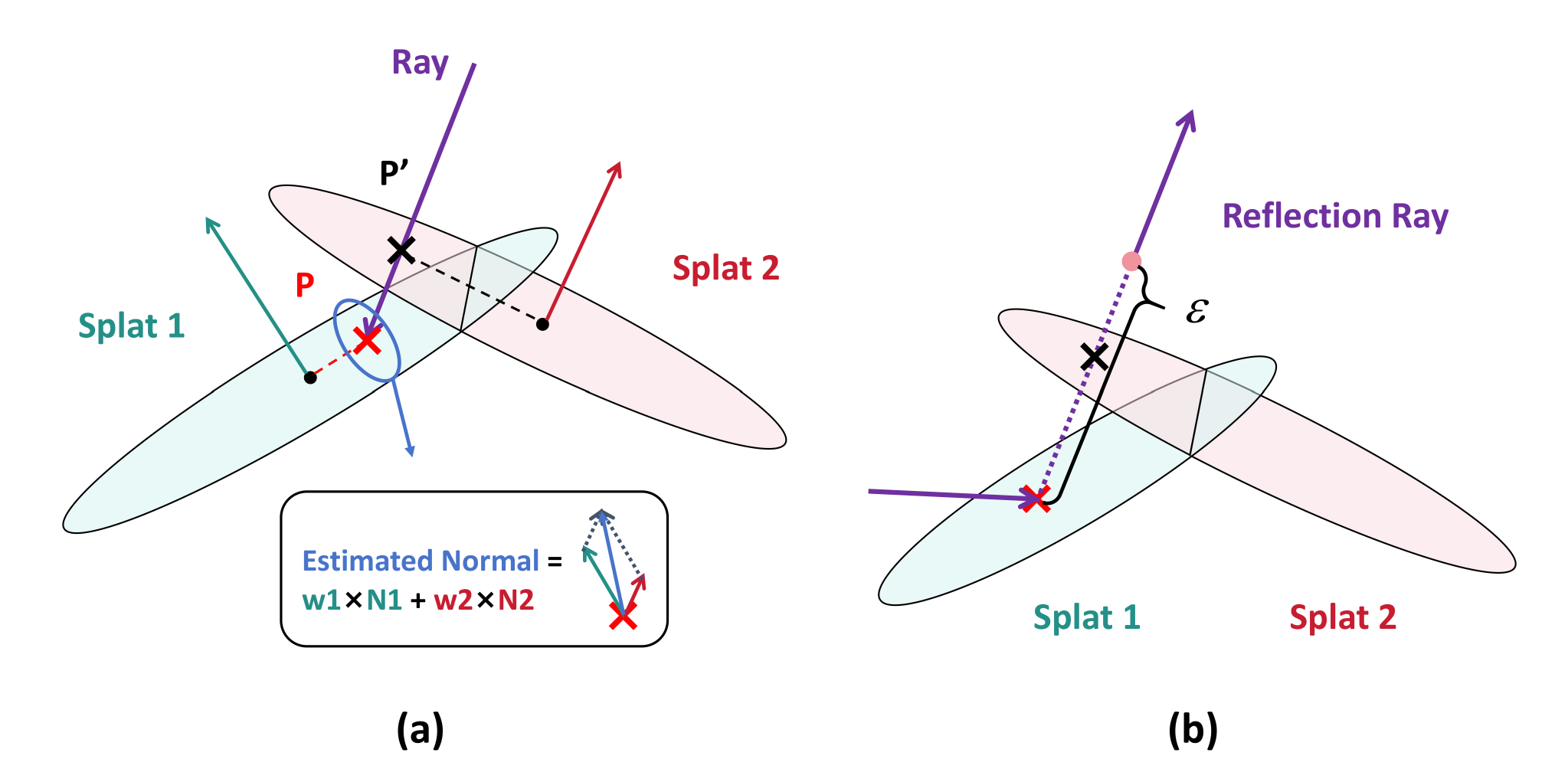}
\caption{When dealing with ray-splat intersections, additional testing becomes essential. In Figure (a), two splats overlap, resulting in the ideal intersection point $\mathbf{P}$ rather than the closer one. Moreover, a blending technique is employed to achieve a smoother normal estimation. In Figure (b), shifting the origin of the secondary ray aids in avoiding erroneous intersections.}
\label{fig_4}
\end{figure}

The execution process of our CFB-assisted MBC is outlined in Algorithm \ref{alg:alg3}. To initiate MBC, we first generate the required GFBs using the method introduced in the previous subsection. Due to the neural network's inability to recover discontinuous signals, the edge portions of a GFB often contain unreliable values. Consequently, we detect and filter these edges using a Canny operator to eliminate error values. Subsequently, we project the GFBs onto $256\times 256\times 256$ grids and fuse the geometrical information by averaging the values within each grid. For tracing rays on these GFBs, we create a single splat for each pixel of the frame buffers, where the splat's position and normal vector are determined by a depth and normal map. Unlike creating splats directly on a point cloud, where determining the radius is complex, we can easily determine the radius of the desired splats here. We adopt a strategy to calculate the radius $R$ of a single splat based on the empirical equation \eqref{deqn_ex22}, considering the angle between the projection direction and the normal direction.
\begin{equation}
    \label{deqn_ex22}
    R=\begin{cases}
        \sqrt{2}L\sqrt{\dfrac{||\mathbf{n}||\times||\mathbf{p}||}{|\mathbf{n} \cdot \mathbf{p}|}}, &{\sqrt{\dfrac{||\mathbf{n}||\times||\mathbf{p}||}{|\mathbf{n} \cdot \mathbf{p}|}} < 2.5} \\
        3.535L, &{\sqrt{\dfrac{||\mathbf{n}||\times||\mathbf{p}||}{|\mathbf{n} \cdot \mathbf{p}|}} \geq 2.5}
    \end{cases}
\end{equation}
\noindent Here, $L$ denotes the side length of a pixel, approximately one-256th of the longest edge of the bounding box in our experiments. $\mathbf{n}$ represents the normal vector of a splat, while $\mathbf{p}$ indicates the projection direction vector. To prevent outliers, we set an upper bound for $R$. Following the initialization of the splats, we construct an AS \cite{ref48} for them to expedite ray traversal. 

For ray-splat intersections, additional testing is required due to the potential overlap between splats. We adopt similar strategies as outlined in \cite{ref21}. As shown in Fig. 4(a), the nearest intersection position is located at $\mathbf{P'}$ when using the same criterion as for ray-mesh intersections. However, it is more suitable to place the intersection point at $\mathbf{P}$ because it is closer to the center of Splat 1. In practice, we examine a sphere with radius $r$, centered at the current candidate intersection point, to determine whether there are any more suitable positions. Subsequently, we blend the normal vectors within the sphere based on equation \eqref{deqn_ex23} to achieve a smooth normal vector estimation. 
\begin{equation}
    \label{deqn_ex23}
    \mathbf{n} = \frac{\displaystyle\sum^{K}_{i=1}\left(1-\dfrac{||\mathbf{p}_i - \mathbf{c}_i||}{R_{i}}\right)\mathbf{n}_i}{\displaystyle\sum^{K}_{i=1}\left(1-\dfrac{||\mathbf{p}_i - \mathbf{c}_i||}{R_{i}}\right)}
\end{equation}
\noindent Here, we denote the hit positions on Splat $i$ as $\mathbf{p}_i$, and the center and radius of Splat $i$ as $\mathbf{c}_i$ and $R_i$, respectively. For reflection rays, we introduce a hyperparameter $\epsilon$ to establish the minimum travel distance, thereby preventing early termination at a neighboring overlapping splat, as illustrated in Fig. 4(b). With these considerations, we have now presented all the details of our ray-splat intersection test method. The subsequent steps of ray-splat intersection remain consistent with those of the ray-mesh intersection.  

\subsection{Time Complexity}
The time complexity analysis of PointEMRay can be segmented into two key components: screen-based PRI and GFB-assisted MBC.

Screen-based PRI involves ray tube tracing and neural network inference. The time complexity of ray tube tracing relies heavily on factors such as scene complexity, ray tube structure, and search strategy, making it challenging to estimate precisely. As a coarse approximation, it can be represented as $O(S\text{log}N)$, where $N$ is the number of points and $S$ is a constant integer determined by point cloud density and ray tube dimensions. Considering the Top-$K$ searching operation, the total time complexity of ray tube tracing becomes approximately $O(M_1S\text{log}K\text{log}N)$, where $M_1$ denotes the number of ray tubes. On the other hand, the neural post-processing stage incurs a time complexity of about $O(M_1(L_e + 3L_d))$, where $L_e$ and $L_d$ refer to the number of layers in the encoder and decoder, respectively. Consequently, the overall time complexity of screen-based PRI can be estimated as $O(M_1(L_e+3L_d+S\text{log}K\text{log}N))$.

GFB-assisted MBC encompasses GFB processing, splat ray tracing, and field computation. GFB processing entails a time complexity of roughly $O(PM_1)$, where $P$ denotes the number of GFBs. Splat ray tracing, on the other hand, carries a complexity no greater than $O(BM_2\text{log}(PM_1))$, with $B$ representing the maximum bounce count and $M_2$ indicating the number of rays. Finally, field computation incurs a complexity of $O(M_2)$. Consequently, the total time complexity of GFB-assisted MBC can be approximated as $O(M_2B\text{log}(PM_1) + PM_1)$.

In summary, the overall time complexity of PointEMRay can be expressed as $O(M_2B\text{log}(PM_1) + PM_1(L_e+3L_d+S\text{log}K\text{log}N))$.

\section{Numerical Results}
This section provides the numerical results derived from several experiments to demonstrate the accuracy and efficiency of PointEMRay.

\subsection{Validation of Screen Based PRI}
\begin{figure}
    \centering
    \includegraphics[width=3.5in]{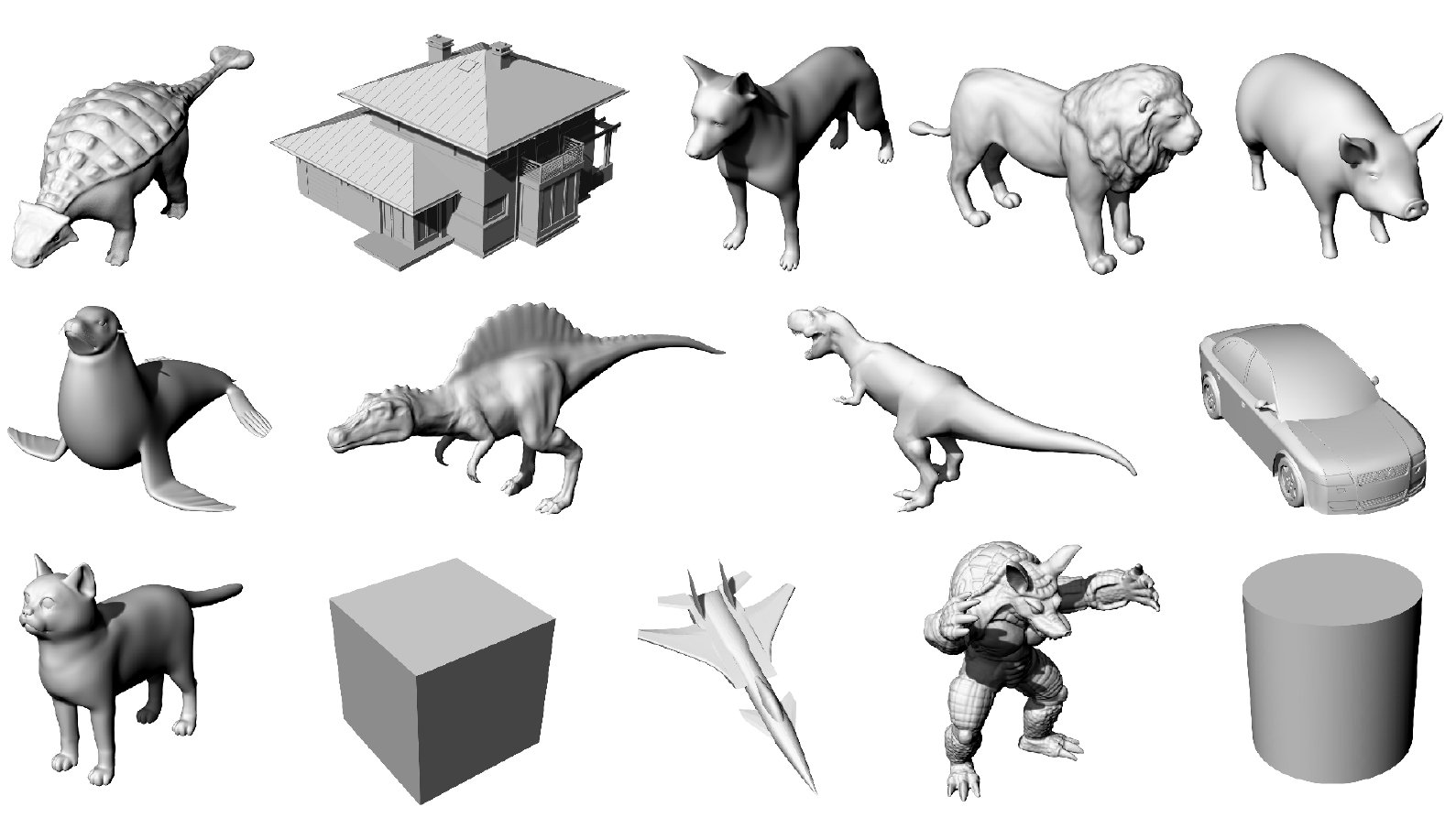}
    \caption{Some of the meshes included in our dataset.}
    \label{fig_5}
\end{figure}

\begin{figure*}[!t]
\centering
\includegraphics[width=7.0in]{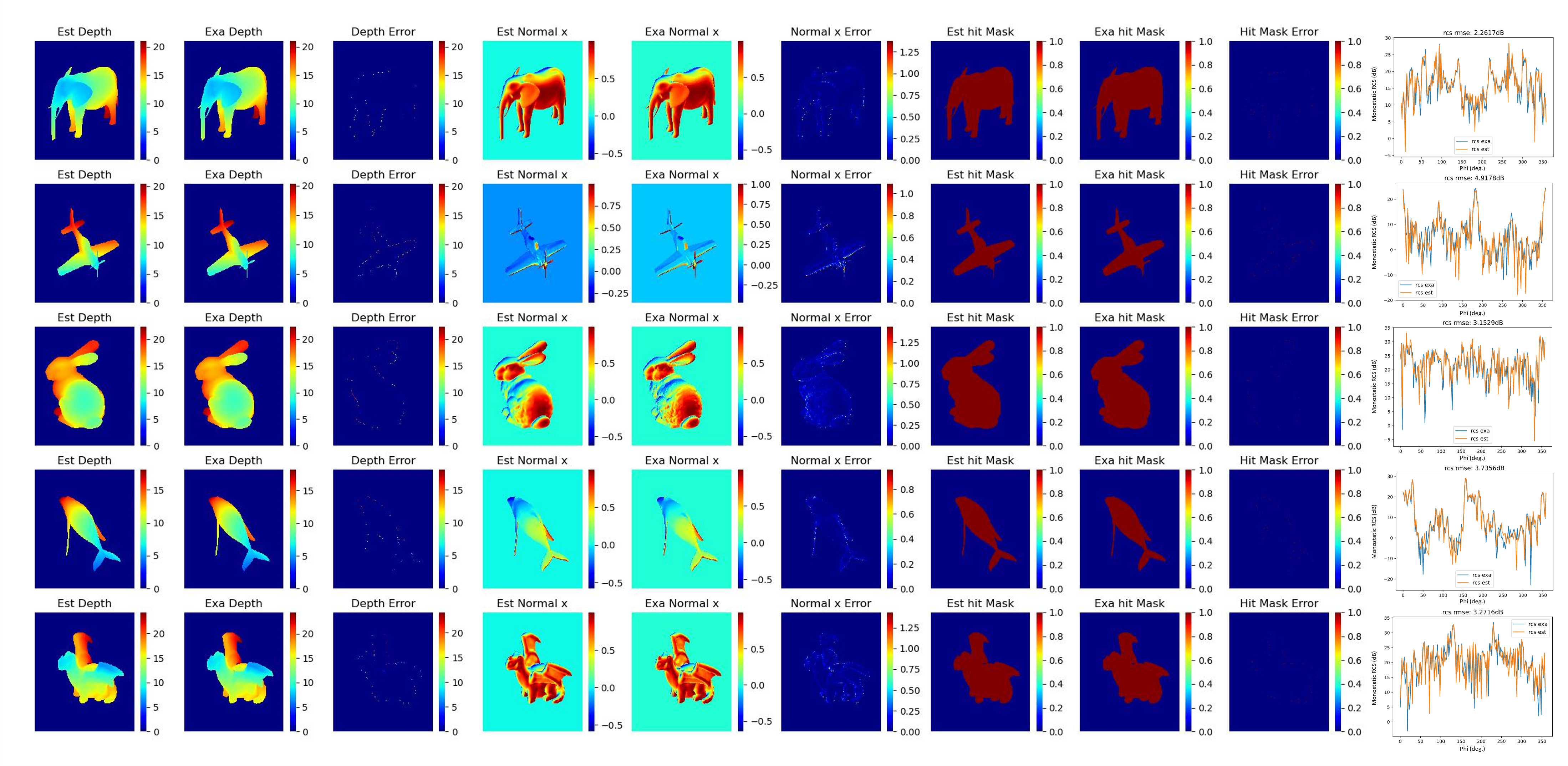}
\caption{Here are the snapshots of prediction results at $\theta=60^\circ$, $\phi=60^\circ$, and monostatic RCS plots of 5 sample models. In columns 1 to 9, you'll find the predicted results, ground truth, and absolute error of depth, normal maps, and intersection masks sequentially. In column 10, you can see the plots of monostatic RCS with computing root mean square errors (RMSEs) of 2.6217dB, 4.9178dB, 3.1529dB, 3.7356dB, and 3.2716dB from top to bottom.}
\label{fig_6}
\end{figure*}

\begin{figure}
    \centering
    \includegraphics[width=3.5in]{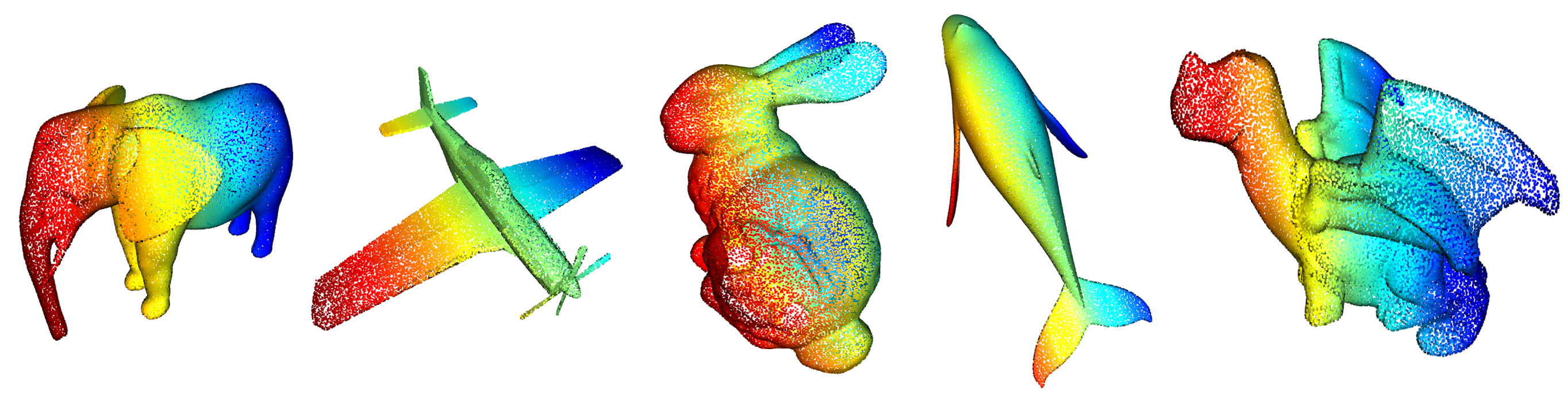}
    \caption{Testing point clouds in Fig. 6.}
    \label{fig_7}
\end{figure}

\begin{figure}
    \centering
    \includegraphics[width=3.5in]{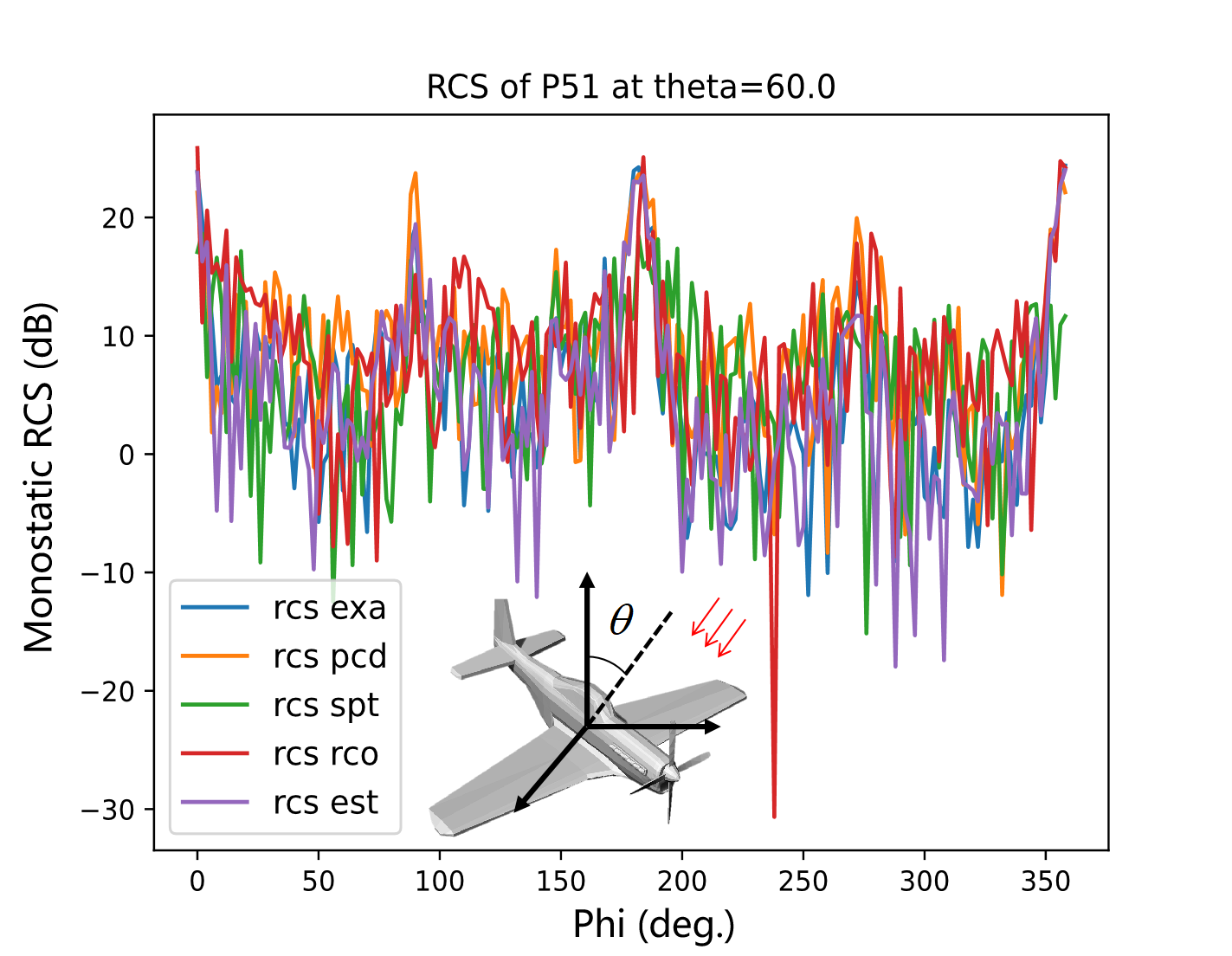}
    \caption{Monostatic RCS of a plane simulated at the angle of $\theta=60^\circ$ with different methods. In the legend, "exa" refers to the ground truth, "pcd" denotes the curve derived from ray tube tracing, "spt" denotes splatting, "rco" denotes Poisson reconstruction, and "est" represents the curve derived from our method.}
    \label{fig_8}
\end{figure} 

\begin{figure*}[!t]
\centering
\includegraphics[width=7.0in]{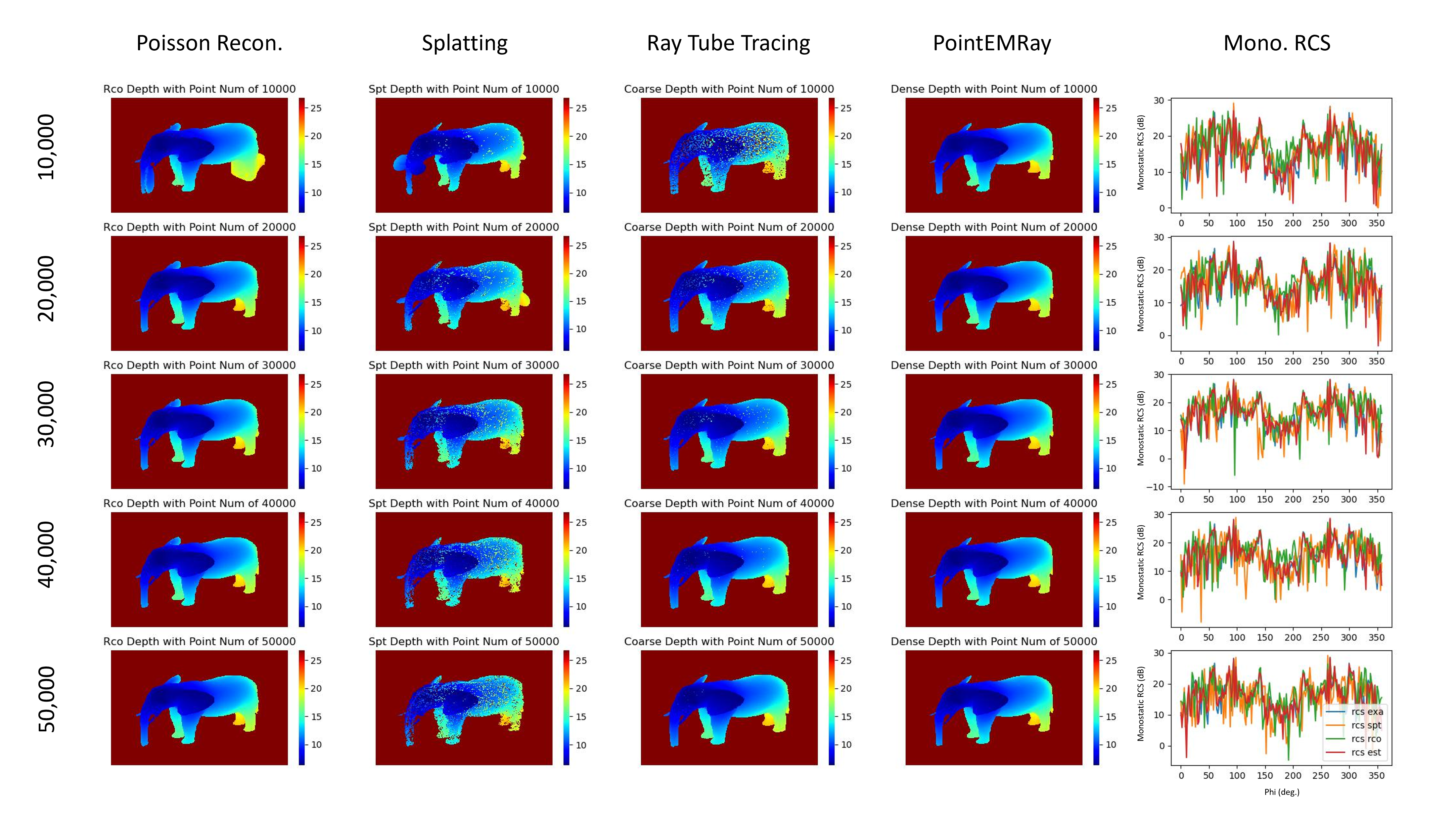}
\caption{The experiment explores the generalization of our method concerning the sampling number. The sampling numbers range gradually from 10k to 50k from top to bottom. Columns 1 to 4 display snapshots of depth maps at $\theta=60^\circ, \phi=60^\circ$ generated by Poisson reconstruction, splatting, ray tube tracing, and our method sequentially. Column 5 showcases the monostatic RCS plots.}
\label{fig_9}
\end{figure*}

To facilitate screen-based PRI, we trained a neural network illustrated in Fig. 3. The dataset utilized for training this network was curated by initially acquiring a series of .obj files from Free3D, accessible at https://free3d.com/. Subsequently, we employed Open3D, an open-source 3D data processing library, to convert these mesh files into point clouds. This dataset comprised 25 meshes, spanning diverse categories such as simple geometries, buildings, vehicles, and animals. Among these, 15 meshes were designated for training, 5 for validation, and 10 for testing. Fig. 5 displays some examples of these meshes.

We observed variations in the sizes of the downloaded meshes, so we uniformly scaled each model such that the longest side of the bounding box measured 20 meters. Following scaling, we initiated the sampling procedure. To ensure the neural network's ability to generalize across point clouds with different densities, we specified sampling numbers ranging from 10k to 50k, with increments of 10k. Additionally, we randomly set the relative radius of the traced ray tubes, varying from 1 to 3 (equivalent to 1.0x to 3.0x the "pixel" edge length, where "pixel" denotes a grid formed by adjacent rays).

In our experiments, the incident plane wave operated at a frequency of 500 MHz, corresponding to a wavelength of 0.6 meters. To define the screen or the plane of the incident wavefront, we set the interval between adjacent rays to one-tenth of the simulation wavelength. With these parameters established, we cast ray tubes onto the point clouds and rays onto the corresponding meshes from 500 different angles for each sample mesh to generate input-output pairs.

The ray tube tracing program was developed in C++ with CUDA. The training process was conducted using PyTorch and executed on an NVIDIA Quadro RTX 8000 GPU. We employed the AdamW optimizer \cite{ref53} during training, with an initial learning rate of 1e-4, decaying by a factor of 0.5 every ten epochs until it reached 1e-6. Ultimately, the neural network was trained for approximately 60 epochs over a span of 3 days.

Here, we present the results of the screen-based PRI experiment. We selected 5 testing meshes and sampled them to generate point clouds, as depicted in Fig. 7. Using a sampling number of 50k and a relative radius of 2.0 for the ray tube, we specifically showcase snapshots for $\theta=60^\circ$ and $\phi=60^\circ$, outlined within the second dashed box in Fig. 6. These images can be grouped into four categories. Columns 1 to 3 display the depth prediction results, with column 1 showing ground truth depth maps obtained from mesh-based ray tracing, column 2 representing the predicted depth maps, and column 3 showing the absolute error between the two depth maps. From the third column, it's evident that the predictions closely match the true values, with only minor defects along the edges. Similarly, columns 4 to 6 display the normal prediction results, and columns 7 to 9 display the intersection mask prediction results. For simplicity, we only show the X-axis component of a normal vector. In both cases of normal vectors and intersection masks, we observe good matches between predictions and ground truth. To further demonstrate the performance of our method, we calculated the monostatic RCS of these selected models based on the prediction results using the PO method. Finally, we plotted the RCS curves in column 10. From the images, we can observe the close match between the curves derived from our method and those obtained from mesh-based ray tracing. 

In addition, we conducted a comparison between our method and other traditional PRI solutions. Employing the same parameters as before, with the incident EM wave operating at 500MHz and being directed at $\theta=60^\circ$, we evaluated the monostatic RCS using two alternative methods: Poisson reconstruction \cite{ref18} and splatting \cite{ref21}. Additionally, we considered ray tube tracing, which provides coarse depth maps as output. For surface normal vector estimation in ray tube tracing, we simply employed principal component analysis (PCA) for each point. In Fig. 8, we present the calculation results for a plane, including the curve of ground truth generated from mesh-based ray tracing and the curves obtained from the three point-based methods. To highlight the distinctions, we computed the root mean square errors (RMSEs) for the aforementioned methods. For the plane model in Fig. 8, the RMSEs of our method, Poisson reconstruction, splatting, and ray tube tracing are 4.9178dB, 8.2061dB, 8.3294dB, and 6.3821dB, respectively. The RMSE results indicate that our method achieves the highest accuracy among these four methods.

We then explored the impact of varying sampling numbers. Selecting an elephant from the testing meshes as our simulation target, in Fig. 9, the sampling numbers range from 10k to 50k with intervals of 10k, while the relative radius of ray tubes remains unchanged at 2. We also included the results of Poisson reconstruction and splatting for comparison. Column 1 in Fig. 9 displays depth maps generated from ray tracing on reconstructed meshes at $\theta=60^\circ$ and $\phi=60^\circ$; column 2 represents depth maps from splatting; column 3 depicts results from ray tube tracing; column 4 showcases the predicted dense depth maps, and column 5 illustrates the computed monostatic RCS results using different methods. It's evident from Fig. 9 that our method consistently delivers high-quality results across varying sampling numbers. Furthermore, Table I presents the RMSEs of monostatic RCS calculated with different methods. As expected, our method achieves the highest accuracy among all tested candidates, underscoring its robust generalization across different sampling numbers.

\begin{table}[!t]
\caption{Monostatic RCS RMSEs for different methods in Fig. 9}
\label{tab:table1}
\centering
\begin{tabular}{c|c|c|c}
\hline \textbf{Sampling Num.} & \textbf{Poisson (dB)} & \textbf{Splatting (dB)} & \textbf{PointEMRay (dB)} \\
\hline 10,000 & 4.8512 & 5.1061 & \textbf{3.6116} \\
 20,000 & 5.4166 & 5.6578 & \textbf{2.6759} \\
 30,000 & 4.9758 & 6.2078 & \textbf{2.7893} \\
 40,000 & 5.1639 & 6.5209 & \textbf{2.1083} \\
 50,000 & 4.9349 & 5.9702 & \textbf{2.2617} \\
\hline
\end{tabular}
\end{table}

\begin{figure*}[!t]
\centering
\includegraphics[width=7.0in]{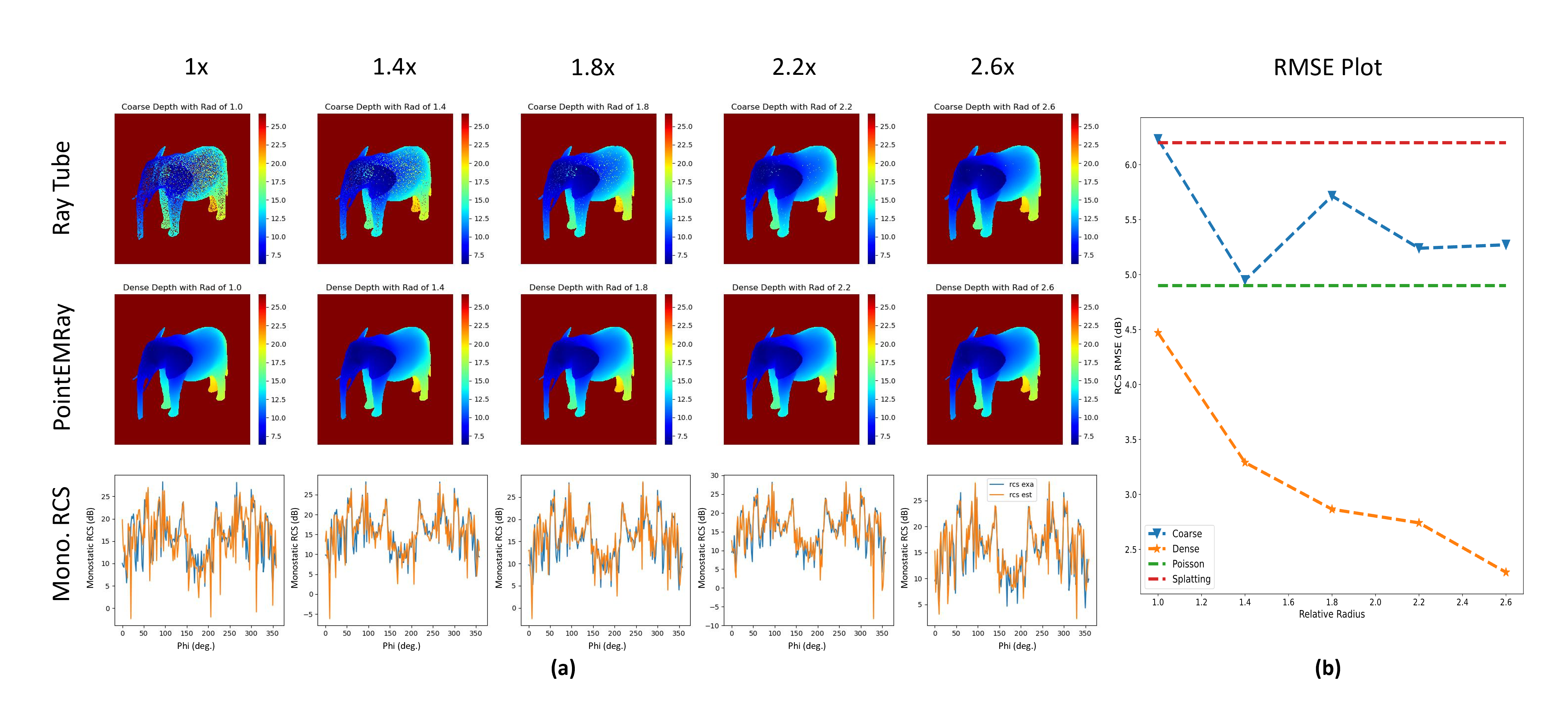}
\caption{An experiment to assess the generalization of our method concerning the relative radius of a ray tube. In (a), Rows 1 and 2 display snapshots of depth maps at $\theta=60^\circ$ and $\phi=60^\circ$ generated by ray tube tracing and our method, respectively, with the relative radius varying from 1 to 2.6. Row 3 presents the corresponding plots of monostatic RCS. In (b), we recorded the computing RMSEs of different methods with dashed lines.}
\label{fig_10}
\end{figure*}

Another critical hyper-parameter, the radius of ray tubes, may also influence the prediction results. To examine this effect, we conducted an experiment testing the response of our neural network to various relative radius values. In Fig. 10(a), the results of this experiment are presented. Similarly, snapshots of depth maps at $\theta=60^\circ$ and $\phi=60^\circ$ with relative radii ranging from 1 to 2.6 are displayed in the first two rows of Fig. 9(a). Row 1 shows the depth maps generated by ray tube tracing, Row 2 presents the results of our method, and Row 3 showcases the calculated monostatic RCS curves based on both coarse and dense geometry at $\theta=60^\circ$. From Fig. 9(a), it is evident that the dense depth map remains of high quality regardless of the change in relative radius. Further insights can be gained from the RMSE plots. In Fig. 9(b), we observe the lines of RMSE changing with relative radius, noting that our method consistently achieves very high accuracy (exhibiting the lowest RMSE) within the varying interval of relative radius. These results underscore the robustness and generalization capability of our method concerning the radius of ray tubes.

\subsection{Validation of GFB Assisted MBC}
\begin{figure}
    \centering
    \includegraphics[width=3.5in]{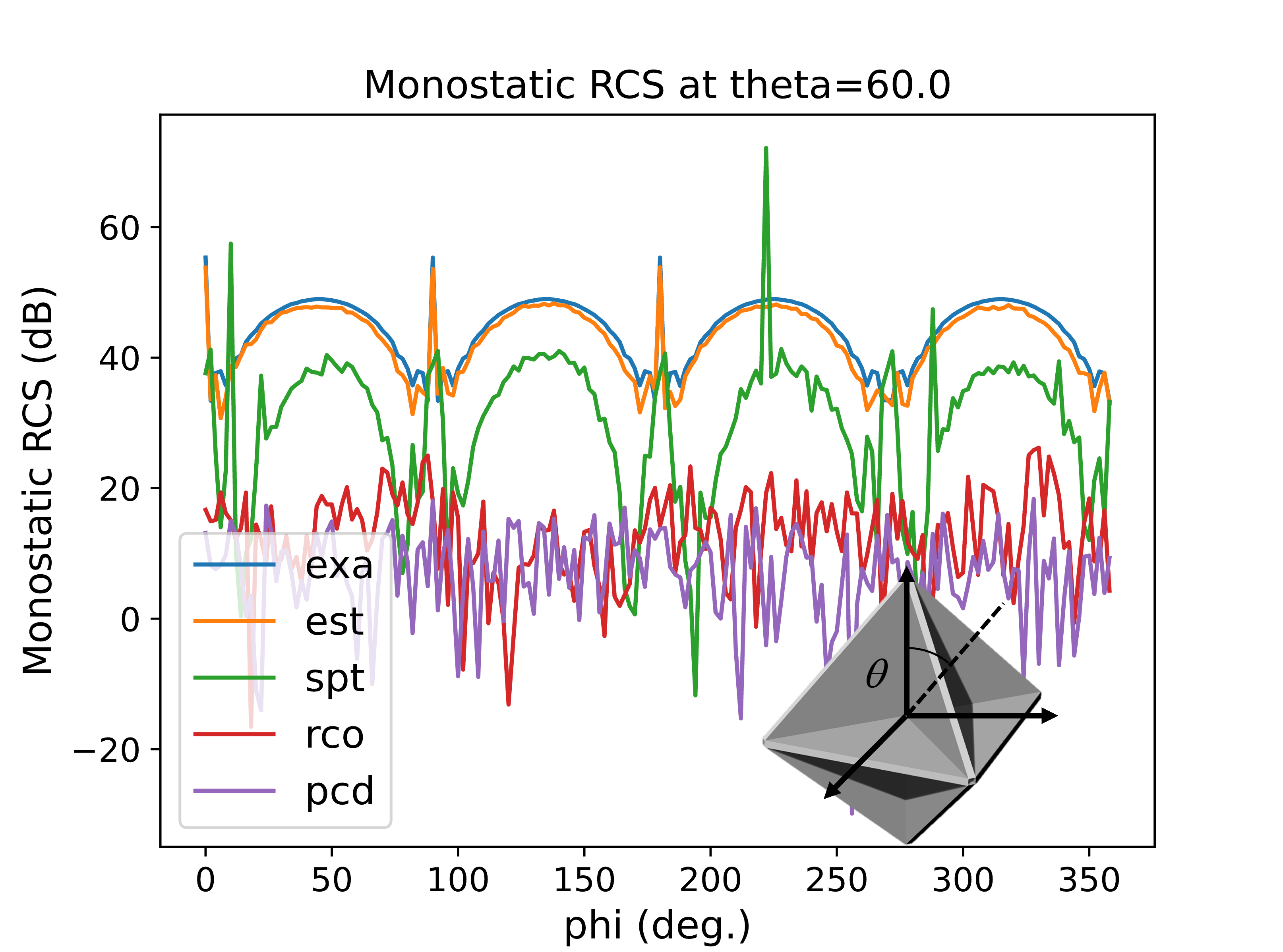}
    \caption{Monostatic RCS of an OctAR simulated at the angle of $\theta=60^\circ$ with different methods. In the legend, "exa" refers to the ground truth, "pcd" denotes the curve derived from ray tube tracing, "spt" denotes splatting, "rco" denotes Poisson reconstruction, and "est" represents the curve derived from our method.}
    \label{fig_11}
\end{figure}

To highlight the multipath effects and demonstrate the benefits of our GFB-assisted MBC, we've developed a unique test model called OctAR, consisting of 8 trihedral angle reflectors, as depicted in Fig. 11. Unlike typical real-world objects, OctAR exhibits exceptionally strong reflections in all directions when exposed to incident EM waves. As a result, accurately simulating the RCS of OctAR necessitates a sophisticated MBC algorithm capable of delivering high-quality performance. Despite its seemingly straightforward geometry, reconstructing OctAR from point cloud data poses significant challenges due to its numerous plate and corner structures \cite{ref33}. Considering these factors, we've chosen OctAR as the target object for our subsequent experiments. To expedite the computational process, we've employed GPU parallel computation techniques and coded using C++ and CUDA.

We began by computing the monostatic RCS of OctAR at $\theta=60^\circ$. Following our previous procedures, we operated the incident EM field at 500MHz and scaled the OctAR model to ensure the longest edge of its bounding box measured 20 meters. Employing a sampling number of 50k and setting the relative radius of a ray tube to 2, we collected predicted GFBs from 8 observation angles ($\theta=60^\circ, 120^\circ$ and $\phi=45^\circ, 135^\circ, 225^\circ, 315^\circ$) to reconstruct the OctAR model. After preprocessing the data, which included edge removal, and fusion, we obtained approximately 200k splats representing the dense geometry of OctAR, quadrupling the original number of points. Subsequently, our MBC algorithm was executed to derive the simulated monostatic RCS curve. In comparison, we applied Poisson reconstruction, the splatting method, and ray tube tracing to perform the same task. Fig. 11 illustrates the simulated results, showcasing the impressive accuracy of our method, which achieved the best alignment with ground truth among all tested methods. While Poisson reconstruction and ray tube tracing failed to provide reasonable simulations, splatting exhibited more precision but still showed significant deviations. Further analysis revealed RMSEs for different methods: 1.4665dB for our method, 31.5443dB for Poisson reconstruction, 15.3948dB for splatting, and 37.2962dB for ray tube tracing, consistent with our observations.

\begin{figure*}[!t]
\centering
\includegraphics[width=7.0in]{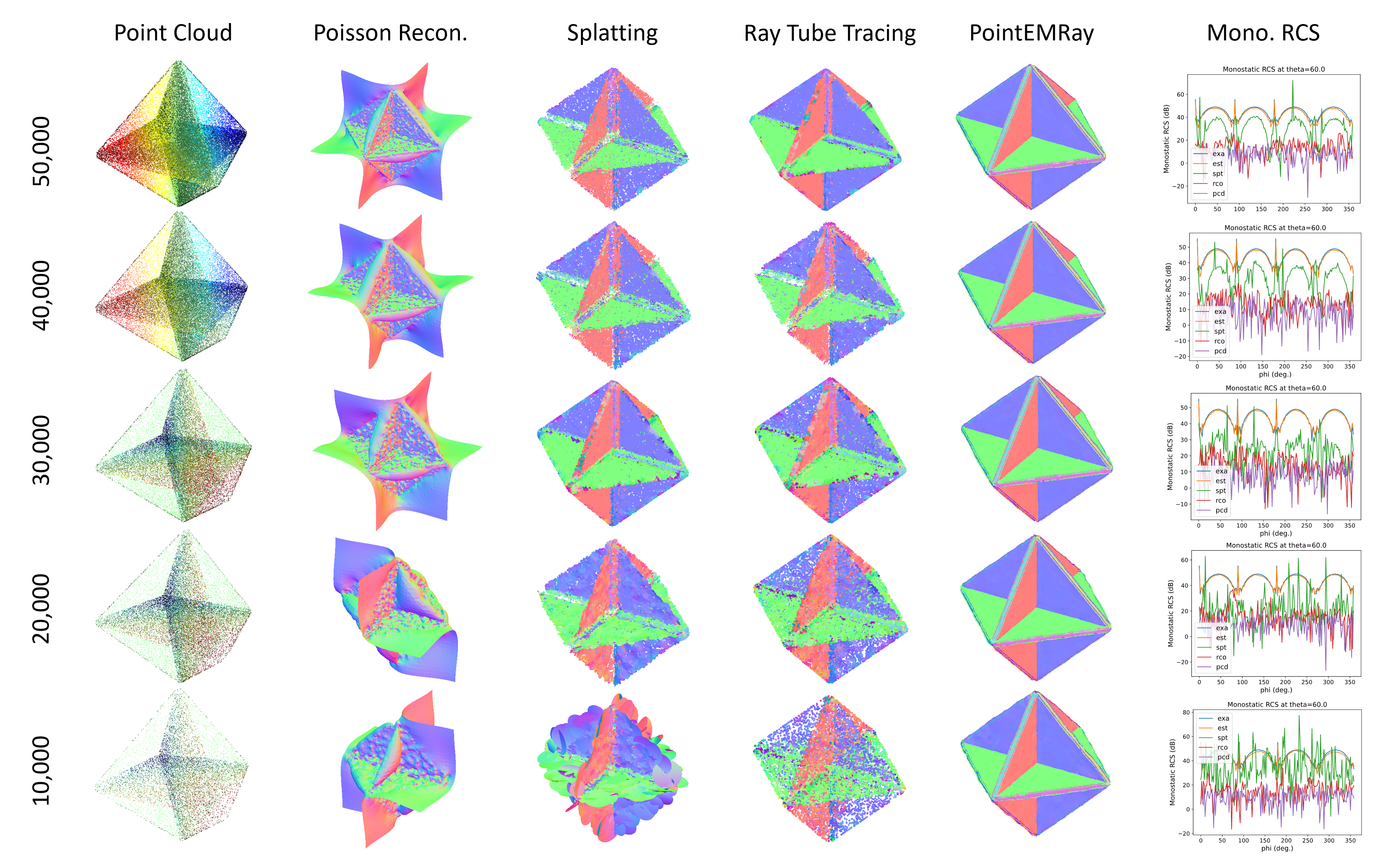}
\caption{The experiment explores the generalization of our MBC algorithm concerning the sampling number. The sampling numbers range gradually from 10k to 50k from bottom to top. Column 1 shows the evaluated point cloud of an OctAR with different sampling numbers. Columns 2 to 5 display snapshots of normal maps at $\theta=60^\circ, \phi=30^\circ$ generated by Poisson reconstruction, splatting, ray tube tracing, and our method sequentially. Column 6 showcases the monostatic RCS plots.}
\label{fig_12}
\end{figure*}

Next, we varied the sampling number to assess the robustness of our method. Maintaining the relative radius at 2 and using the same EM field configuration as before, we opted to showcase the computing results using normal maps instead, as they offer clearer geometry details. Fig. 12 presents the outcomes of our experiment. Column 1 displays snapshots of the input point cloud, where the sparsity of geometry gradually increases with decreasing sampling numbers. Columns 2 to 5 exhibit snapshots of normal maps at $\theta=60^\circ, \phi=30^\circ$ generated by single ray tracing (ray tubes). Notably, our method consistently recovers the most accurate geometry, demonstrating minimal sensitivity to changes in the sampling number. In contrast, Poisson reconstruction exhibits the poorest performance, failing to capture even the rough outline of OctAR regardless of the sampling number. Splatting, while superior to Poisson reconstruction, shows gradual improvement in geometry as the sampling number increases. Similar observations apply to ray tube tracing, as depicted in Fig. 12. Column 6 depicts the monostatic RCS plots, accompanied by RMSE values for different methods listed in Table II. As expected, our method yields the lowest RMSE, underscoring its outstanding performance. However, an intriguing finding emerges: despite providing a passable recovery of geometry, ray tube tracing yields the largest RMSE. Further experimentation led us to attribute this phenomenon to the limited capability of ray tube tracing to simulate multipath effects accurately, as ray tubes often intersect the point cloud in close proximity to the reflection or refraction position. 

\begin{table}[!t]
\caption{Monostatic RCS RMSEs for different methods in Fig. 12}
\label{tab:table1}
\centering
\begin{tabular}{c|c|c|c|c}
\hline \textbf{SN} & \textbf{PR (dB)} & \textbf{SPT (dB)} & \textbf{RTT (dB)} & \textbf{PointEMRay (dB)} \\
\hline 10,000 & 29.9882 & 14.9385 & 42.7496 & \textbf{2.5713} \\
 20,000 & 29.4161 & 24.2103 & 40.9720 & \textbf{1.8264} \\
 30,000 & 30.5043 & 21.2963 & 40.6381 & \textbf{1.9161} \\
 40,000 & 31.3127 & 15.4488 & 39.9015 & \textbf{1.9061} \\
 50,000 & 31.5443 & 15.9348 & 37.2962 & \textbf{1.4665} \\
\hline
\end{tabular}
\end{table}

Finally, we delved into the impact of the ray tube radius, a crucial parameter for both ray tube tracing and our method. With the sampling number set to 30k, we generated corresponding normal maps and RCS plots, as illustrated in Fig. 13. In the first two rows, the normal map of ray tube tracing gradually refines with the increase in relative radius, while our method consistently maintains high-quality results. This suggests that our method is robust to changes in the radius of a ray tube. Similar trends are observed in Row 3, where the orange curve (our method) consistently aligns well with the blue curve (ground truth), exhibiting RMSEs of 2.4362 dB, 1.9949 dB, 1.9315 dB, 1.9900 dB, and 1.9804 dB from left to right. Conversely, the green curve (ray tube tracing) is notably influenced by the radius of a ray tube, evident in the series of plots. Thus, we conclude that our method demonstrates strong generalization across various ray tube radii.

\begin{figure*}[!t]
\centering
\includegraphics[width=7.0in]{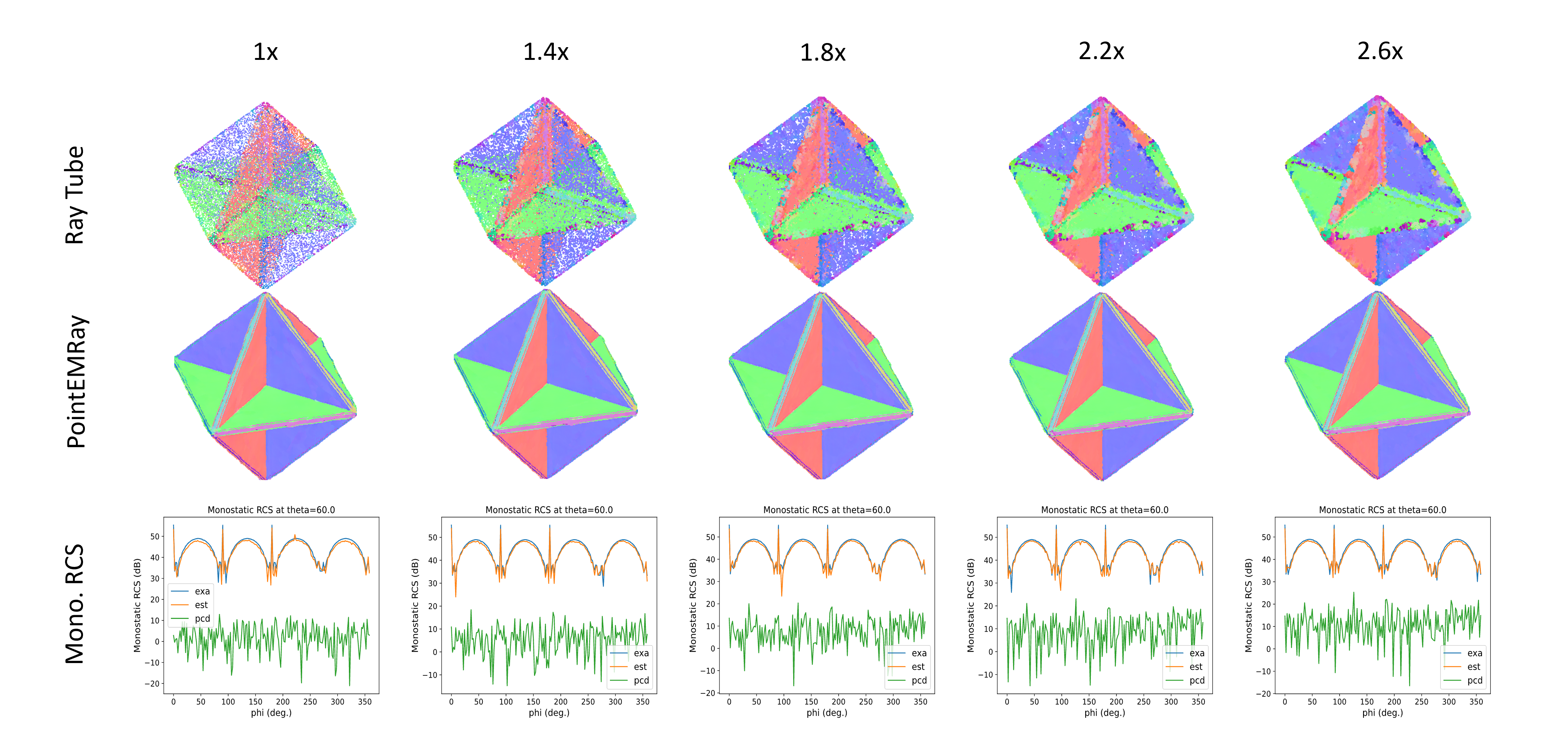}
\caption{The experiment to examine the effect of the relative radius of a ray tube. Rows 1 and 2 display snapshots of normal maps at $\theta=60^\circ$ and $\phi=30^\circ$ generated by ray tube tracing and our method, respectively, with the relative radius varying from 1 to 2.6. Row 3 presents the corresponding plots of monostatic RCS.}
\label{fig_13}
\end{figure*}

\subsection{Efficiency Test}
Discussion of the efficiency of PointEMRay will be divided into two parts: screen-based PRI and GFB-assisted MBC.

Based on the experiments conducted, it takes approximately 0.025 seconds to generate a typical size (height and width about $50\lambda$) coarse depth map through ray tube tracing. Following this, the average inference time of our neural network is approximately 0.05 seconds, as it processes such a typical size input depth map to generate the corresponding GFBs. According to the estimations made in Section III, the execution time of screen-based PRI will grow linearly as the size of the screen increases. Therefore, the total time to complete screen-based PRI for a single observation angle will be approximately 0.075 seconds. Our method still has room for acceleration; for example, we have not leveraged the power of batches when using the neural network.

For GFB-assisted MBC, the time consumption is comparable to traditional ray tracing algorithms. We recorded the execution time of the 3-bounce MBC procedure when computing the monostatic RCS for an OctAR model. The average MBC time is about 0.03 seconds. Additionally, we recorded the average EM field computing time, which is less than 1 millisecond. Thus, the average total time to simulate the monostatic RCS of an OctAR for a single angle is no more than 0.04 seconds, which actually researches real-time level. However, that is still not the key efficiency point of our PointEMRay. To recover the whole geometry, we will execute several times the PRI procedure. Taking the OctAR (sampling number equals 50k) as an example: the total PRI time is about 0.4 seconds, and the data preprocessing time is about 2.4371 seconds. That is to say, we only spend about 3 seconds to recover the geometry from a 50k point cloud, reaching a speed of about 0.6 seconds per 10k points, which far exceeds the records found in previous literature \cite{ref25, ref33, ref34}! Based on these data, we can demonstrate the efficiency of our PointEMRay.

\section{Conclusion}
In this paper, we introduce a novel framework called PointEMRay, designed to conduct traditional SBR simulations on point-based geometry. To address this goal, we tackle two main challenges: PRI and MBC. For PRI, we propose a screen-based approach inspired by SPRIM, leveraging deep learning for efficiency. Initially, we obtain coarse depth maps through ray tube tracing, followed by the generation of dense depth maps, normal maps, and intersection masks using a neural network. These generated maps are collectively referred to as GFBs. For MBC, we draw inspiration from SLAM techniques and introduce GFB-assisted MBC. Specifically, we aggregate GFBs obtained from various observation angles to reconstruct the complete geometry. Subsequently, we execute ray tracing algorithms on these GFBs and compute the scattering EM field.

To validate the accuracy and efficiency of PointEMRay, we conduct several numerical experiments in this paper. Our results demonstrate the superior performance of PointEMRay, potentially surpassing traditional methods such as Poisson reconstruction and splatting, particularly in terms of accuracy. 

Despite the promising outcomes of our research, several limitations remain to be addressed. For instance, we struggle to effectively restore fine structures like surface textures. Moreover, the MBC phase often involves generating and tracking a large number of splats, which can significantly impact solution efficiency. Additionally, our current approach only considers the far-field assumption, and introducing distance may introduce further complexities. Looking ahead, we aim to extend our research to real point cloud data and progress from target-level simulations to scene-level simulations.

\section*{Acknowledgments}
This research was carried out at the Institute of Electromagnetic Space, and the State Key Laboratory of Millimeter Wave of Southeast University.



 


\vspace{-33pt}
\begin{IEEEbiography}[{\includegraphics[width=1in,height=1.25in,clip,keepaspectratio]{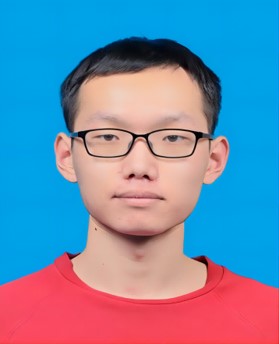}}]{Kaiqiao Yang}
was born in Xi'an, Shaanxi, China, in 1999. He received the B.S. degree in communication engineering from Beijing University of Posts and Telecommunications in 2022. He is currently pursuing the Ph.D. degree in electrical science and technology at Southeast University.

His research interests include the theory of high-frequency asymptotic methods; graphics and intelligent techniques in electromagnetic computing, and the mechanisms of radio wave propagation.
\end{IEEEbiography}

\begin{IEEEbiography}[{\includegraphics[width=1in,height=1.25in,clip,keepaspectratio]{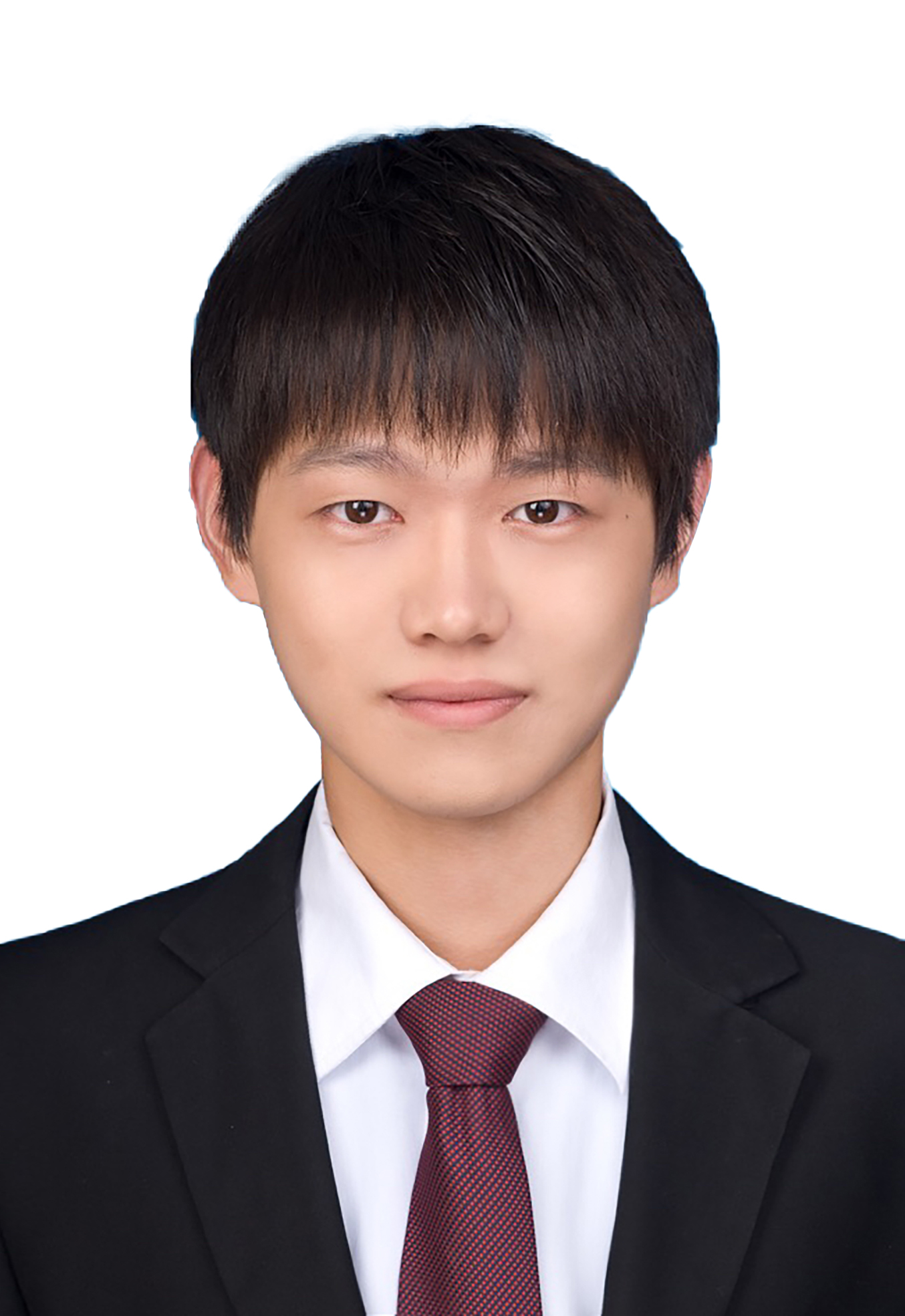}}]{Che Liu}
(Member, IEEE) was born in Suzhou, Jiangsu, China, in 1993. He received the B.Eng. degree in information science and technology and the Ph.D. degree from Southeast University, Nanjing, China, in 2015 and 2022, respectively. He is currently a Zhishan Postdoctor with Southeast University. 

His research interests include computational electromagnetic, meta-material, and deep learning. He is committed to use artificial intelligence technology solving electromagnetic issues, including ISAR imaging, holographic imaging, inverse scattering imaging, automatic antenna design, and diffraction neural network.
\end{IEEEbiography}

\begin{IEEEbiography}[{\includegraphics[width=1in,height=1.25in,clip,keepaspectratio]{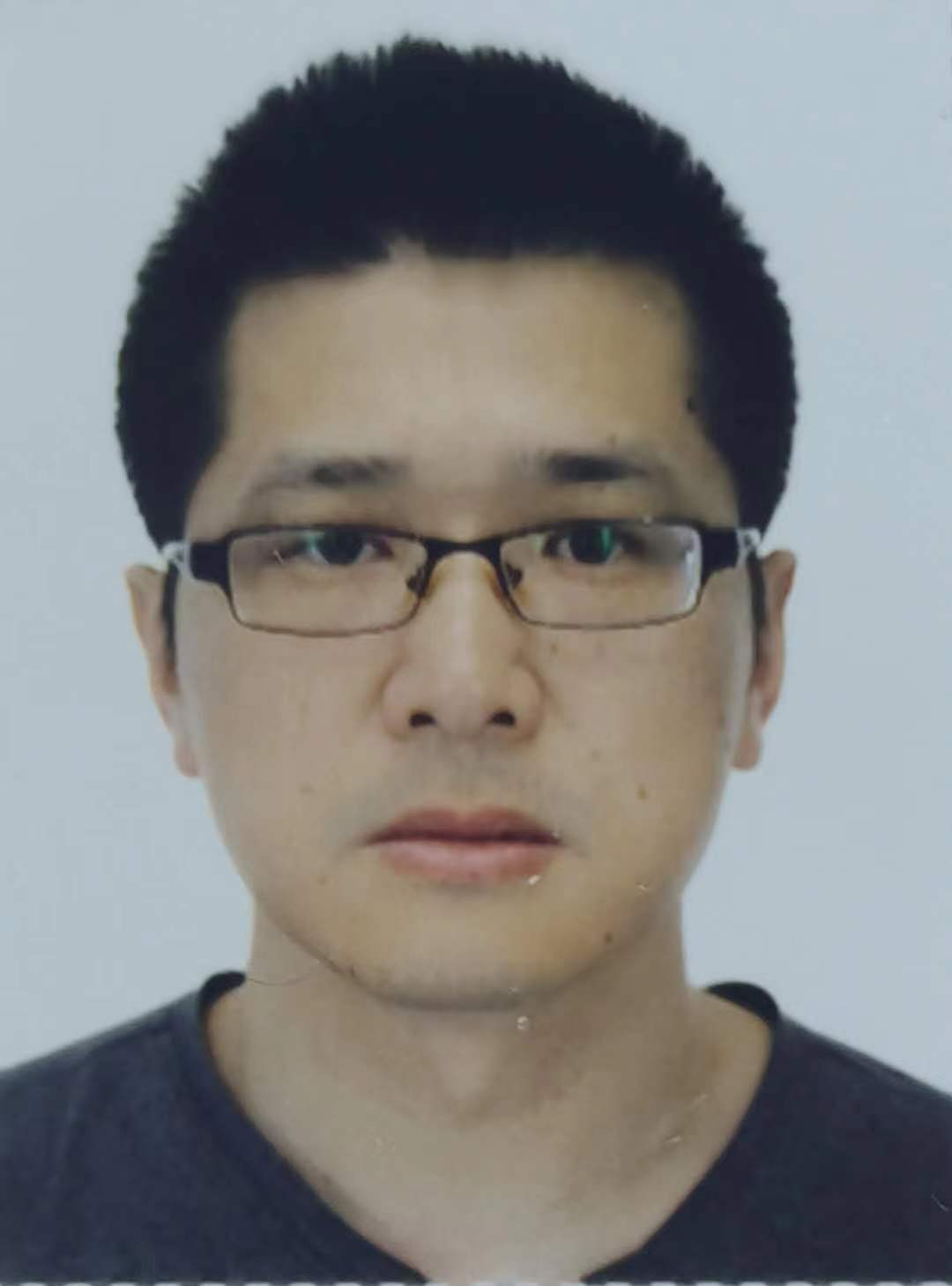}}]{Wenming Yu}
was born in Zhuji, Zhejiang, China, in 1980. He received the B.Sc. and Ph.D. degrees from the Nanjing University of Science and Technology, Nanjing, China, in 2002 and 2007, respectively.

He is a Lecturer with the School of Information Science and Engineering, Southeast University, Nanjing. His research interest is computational electromagnetics.
\end{IEEEbiography}

\begin{IEEEbiography}[{\includegraphics[width=1in,height=1.25in,clip,keepaspectratio]{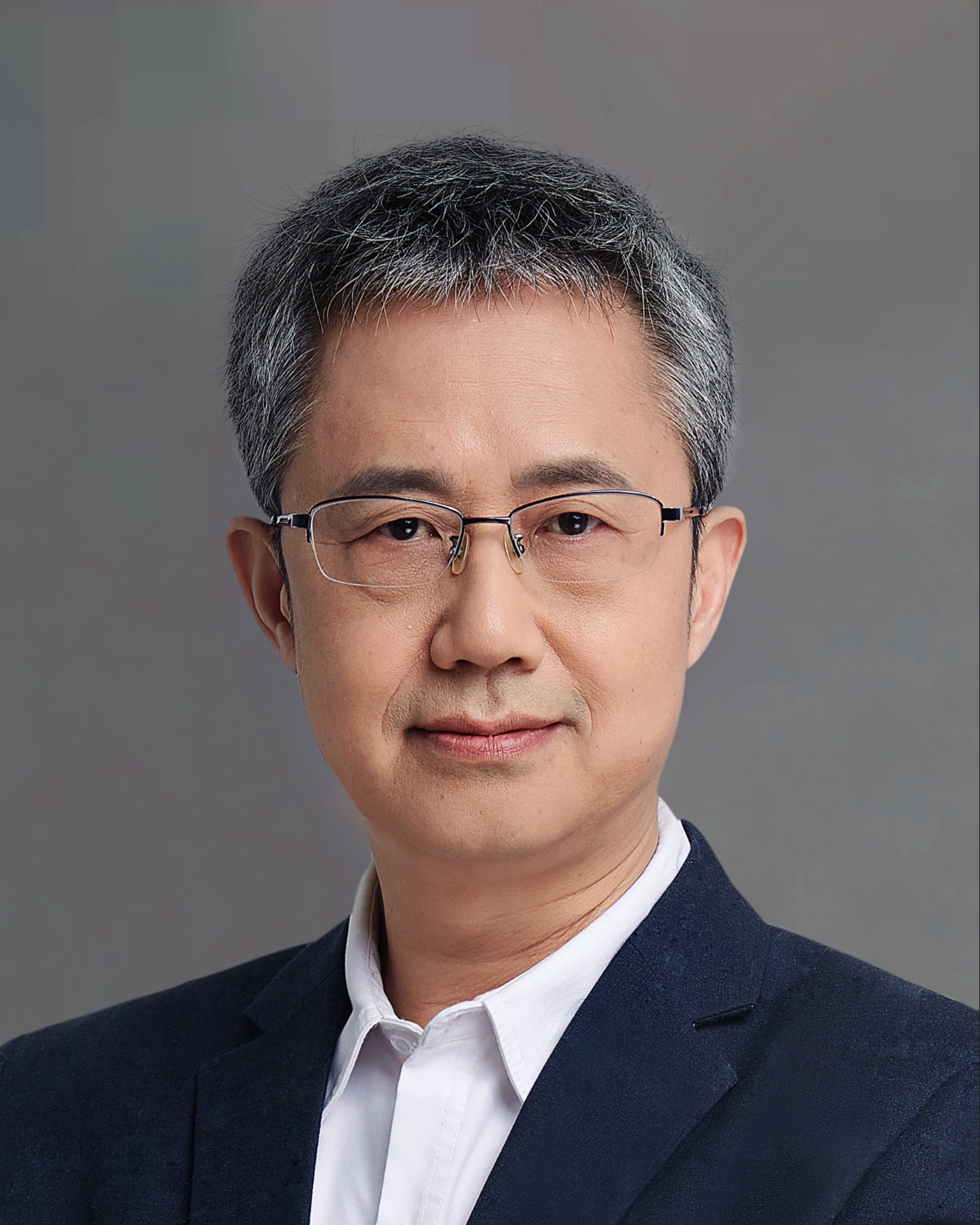}}]{Tie Jun Cui}
(Fellow, IEEE) received the B.Sc., M.Sc., and Ph.D. degrees from Xidian University, Xi’an, China, in 1987, 1990, and 1993, respectively. In March 1993, he joined the Department of Electromagnetic Engineering, Xidian University, and was promoted to an Associate Professor in November 1993. From 1995 to 1997, he was a Research Fellow with the Institut fur Hochstfrequenztechnik und Elektronik (IHE), University of Karlsruhe, Karlsruhe, Germany. In July 1997, he joined the Center for Computational Electromagnetics, Department of Electrical and Computer Engineering, University of Illinois at Urbana-Champaign, Champaign, IL, USA, first as a Postdoctoral Research Associate and then as a Research Scientist. In September 2001, he was a Cheung-Kong Professor with the Department of Radio Engineering, Southeast University, Nanjing, China. In January 2018, he became the Chief Professor of Southeast University. He is an Academician of the Chinese Academy of Science. He is the first author of the books \textit{Metamaterials: Theory, Design, and Applications} (Springer, November 2009), \textit{Metamaterials: Beyond Crystals, Noncrystals, and Quasicrystals} (CRC Press, March 2016), and \textit{Information Metamaterials} (Cambridge University Press, 2021). He has authored or coauthored more than 600 peer-reviewed journal articles, which have been cited by more than 62,000 times (H-Factor 122), and licensed more than 150 patents. His research has been selected as one of the most exciting peer-reviewed optics research Optics in 2016 by Optics and Photonics News Magazine, ten Breakthroughs of China Science in 2010, and many Research Highlights in a series of journals. His work has been widely reported by \textit{Nature News}, \textit{MIT Technology Review}, \textit{Scientific American}, \textit{Discover}, and \textit{New Scientists}. He was the recipient of the Research Fellowship from Alexander von Humboldt Foundation, Bonn, Germany, in 1995, Young Scientist Award from the International Union of Radio Science in 1999.
\end{IEEEbiography}



\vfill

\end{document}